\documentclass[aps,prb,twocolumn,superscriptaddress]{revtex4-2}

\usepackage{graphicx}
\usepackage{amsmath}
\usepackage{ulem}
\usepackage{xcolor}
\usepackage{multirow}

\usepackage{scalerel,amssymb}
\def\msquare{\mathord{\scalerel*{\Box}{gX}}}

\begin{document}

	\title{Spin-photon coupling for atomic qubit devices in silicon}

	\author{Edyta N. Osika}
	\affiliation{School of Physics, University of New South Wales, Sydney, NSW 2052, Australia}
	\affiliation{Silicon Quantum Computing Pty Ltd., Level 2, Newton Building, UNSW Sydney, Kensington, NSW 2052, Australia}
	\author{Sacha Kocsis}
	\affiliation{School of Physics, University of New South Wales, Sydney, NSW 2052, Australia}
	\affiliation{Silicon Quantum Computing Pty Ltd., Level 2, Newton Building, UNSW Sydney, Kensington, NSW 2052, Australia}
	\author{Yu-Ling Hsueh}
	\affiliation{School of Physics, University of New South Wales, Sydney, NSW 2052, Australia}
	\affiliation{Silicon Quantum Computing Pty Ltd., Level 2, Newton Building, UNSW Sydney, Kensington, NSW 2052, Australia}
	\author{Serajum Monir}
	\affiliation{School of Physics, University of New South Wales, Sydney, NSW 2052, Australia}
	\affiliation{Silicon Quantum Computing Pty Ltd., Level 2, Newton Building, UNSW Sydney, Kensington, NSW 2052, Australia}
	\author{Cassandra Chua}
	\affiliation{School of Physics, University of New South Wales, Sydney, NSW 2052, Australia}
	\affiliation{Silicon Quantum Computing Pty Ltd., Level 2, Newton Building, UNSW Sydney, Kensington, NSW 2052, Australia}
	\author{Hubert Lam}
	\thanks{Current address: Fachrichtung Physik, Universit{\"a}t des Saarlandes, 66123 Saarbr{\"u}cken, Germany}
	\affiliation{Centre for Quantum Computation and Communication Technology, School of Physics, University of New South Wales, Sydney, NSW 2052, Australia}
	\author{Benoit Voisin}
	\affiliation{School of Physics, University of New South Wales, Sydney, NSW 2052, Australia}
	\affiliation{Silicon Quantum Computing Pty Ltd., Level 2, Newton Building, UNSW Sydney, Kensington, NSW 2052, Australia}
	\author{Sven Rogge}
	\affiliation{Centre for Quantum Computation and Communication Technology, School of Physics, University of New South Wales, Sydney, NSW 2052, Australia}
	\author{Rajib Rahman}
	\affiliation{School of Physics, University of New South Wales, Sydney, NSW 2052, Australia}
	\affiliation{Silicon Quantum Computing Pty Ltd., Level 2, Newton Building, UNSW Sydney, Kensington, NSW 2052, Australia}
	
	\date{\today}

	\begin{abstract}
	
	Electrically addressing spin systems is predicted to be a key component in developing scalable semiconductor-based quantum processing architectures, to enable fast spin qubit manipulation and long-distance entanglement via microwave photons. However, single spins have no electric dipole, and therefore a spin-orbit mechanism must be integrated in the qubit design. Here, we propose to couple microwave photons to atomically precise donor spin qubit devices in silicon using the hyperfine interaction intrinsic to donor systems and an electrically-induced spin-orbit coupling. We characterise a one-electron system bound to a tunnel-coupled donor pair (1P-1P) using the tight-binding method, and then estimate the spin-photon coupling achievable under realistic assumptions. We address the recent experiments on double quantum dots (DQDs) in silicon and indicate the differences between DQD and 1P-1P systems. Our analysis shows that it is possible to achieve strong spin-photon coupling in 1P-1P systems in realistic device conditions without the need for an external magnetic field gradient.

	\end{abstract}

	
	\maketitle
	
	\section{INTRODUCTION}

Semiconductor spin qubits have now reached high enough figures of merit to envision error-corrected architectures for quantum information processing~\cite{Fowler2012,Hill2015,OGorman2016,Pica2016,Veldhorst2017}. Such architectures are expected to require both short-range spin coupling mechanisms, achievable via $e.g.$ the exchange interaction~\citep{VeldhorstNature15, Hendrickx2020,Watson2018,He2019_SWAP}, as well as chip-scale coupling mechanisms to allow for control electronics and quantum information transfer~\citep{VandersypenNPJ17}. Superconducting microwave cavities, which are used in the context of circuit quantum electrodynamics~\citep{SchoelkopfNature07, Simmonds07}, are also suitable for coupling spin-based qubits at long distances~\citep{Benito2qubit} as the typical gigahertz microwave cavity resonant frequencies can match the energy scales found in semiconductor-based spin devices~\cite{Hanson2007}.\\
    
Direct spin-photon coupling is challenging due to the small magnetic dipole interaction, usually on the order of 10 to 100 Hz \cite{Schoelkopf2008}, between an electron spin and the typical vacuum fluctuations of a resonant cavity. Instead, an electrical coupling between the two entities is preferable, which is also necessary for the development of electrically-driven spin resonance (EDSR) to increase qubit operation speed and scalability~\cite{Golovach2006,Pioro-Ladriere2008,Nadj-Perge2010,Petersson2012,Salfi2016,Corna2018,Watzinger2018,Hendrickx2020,arxiv1P2P, Felix2021}. 
There are two requirements for creating electrical coupling: a charge dipole can be induced by localizing a spin across more than one quantum dot~\cite{Burkard2020}; and spin-to-charge hybridisation can be induced either through a natural spin-orbit mechanism (intrinsic SO, usually weak for electron systems), or by engineering a spin-orbit mechanism (extrinsic SO). An extrinsic SO mechanism can be created, for example, by using the transverse magnetic field gradient of a micro-magnet~\cite{Burkard2006,PioroNano17,Hu2012}. Coupling spins to photons with an extrinsic SO mechanism has successfully been demonstrated in electrostatically defined quantum dots~\citep{Mi2018, Samkharadze2018,PettaNature19,Landig2018,WallraffNatComms19,Koski2020}, yielding spin-photon coupling rates on the order of 10\,MHz. These rates are sufficient to achieve the strong spin-photon coupling regime -- a milestone where the spin-photon interaction rate exceeds both the qubit decoherence and cavity decay rates. While it has not been realised to date, achieving the strong coupling regime using an intrinsic SO mechanism would be desirable to ease device fabrication and scalability.\\

Electron spins bound to $^{31}$P donors in silicon form a relevant qubit platform for quantum information processing with exceptionally long coherence times, demonstrated on the order of seconds~\citep{Muhonen2014, Watson2017}, and the fastest two-qubit gate operations to date~\citep{He2019_SWAP}. Recent progress on donor-bound spin qubits defined with the atomic precision of scanning probe lithography~\citep{Fuechsle2012, KochNatureNano19} have included double and triple quantum dot devices~\cite{Watson2014,PrasannaPRX18,He2019_SWAP}, where each quantum dot can be made of one or more dopants. Importantly, phosphorus atoms host a nuclear spin~\cite{Muhonen2014,Hile2018}, and the hyperfine interaction between the electron and the nuclear spins is the basis of a proposal for a hybrid spin qubit amenable to long-distance coupling via microwave photons~\citep{Tosi2017}. An electrically induced spin-orbit interaction has also recently been observed in donor qubit devices~\cite{weber2018spin}, but its potential in electrically addressing spins remains to be explored.\\
    
Here, we describe a device architecture that combines both the ability to place dopant atoms with atomic precision in silicon and the use of their intrinsic SO mechanisms to couple spin qubits and microwave photons. 
The spin qubit itself consists of a single electron bound to two tunnel-coupled single $^{31}$P donors, whose tunneling frequency is comparable to the resonant frequency of a nearby superconducting coplanar microwave resonator.
Atomistic methods~\cite{Ahmed2009} are used to determine critical system parameters such as the tunnel coupling $t_c$ and the charge dipole moment $d_c$. We then investigate both the hyperfine (HF)~\cite{Tosi2017,Muhonen2014,Hile2018} and the electrically-induced spin-orbit (EISO)~\cite{weber2018spin} mechanisms to realise the necessary spin-charge hybridisation. An effective Hamiltonian approach is used to estimate the ratio, $g_s/g_c$, between the expected spin-photon and charge-photon couplings in a 1P-1P system. We use finite-element simulations of a cavity's electric field at the qubit position to estimate $g_c$, which we find to be comparable to what has recently been measured in gate-defined DQD systems \cite{Mi2018, Samkharadze2018}. Our results suggest that the strong spin-photon coupling regime is achievable for dopant systems in silicon under realistic qubit coherence and device electric field conditions.

\begin{figure}[t]
		\includegraphics[width=8.6cm]{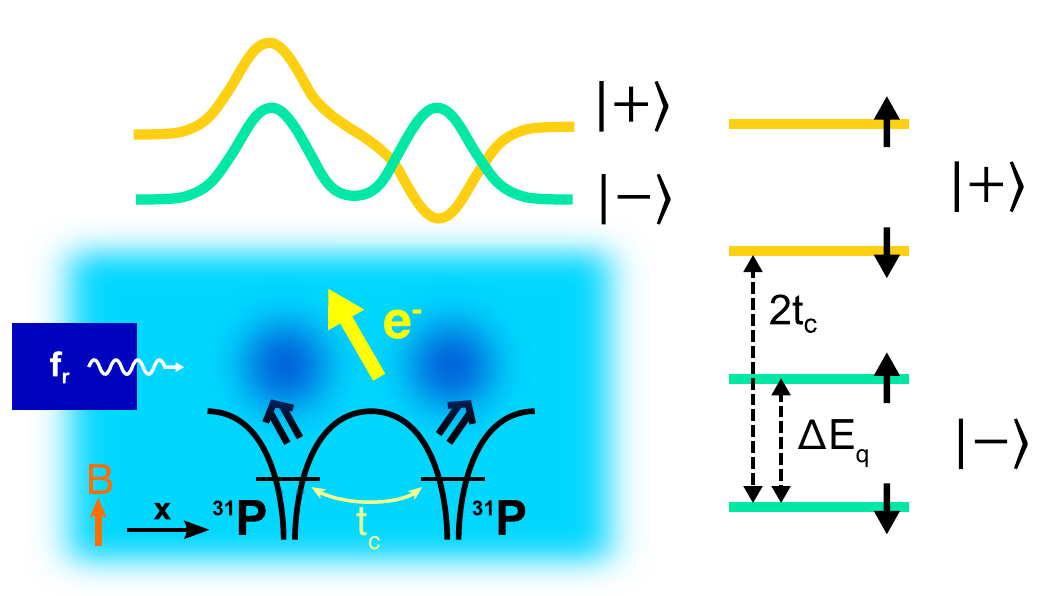}%
	\caption{Proposed device architecture. We consider a single electron (yellow arrow) confined between two phosphorus donors placed in the lattice of bulk silicon and separated in $x$ direction. The two donor states are tunnel coupled ($t_c$) to form the bonding $|-\rangle$ and antibonding $|+\rangle$ orbital states. The qubit is formed between electron spin $\downarrow$ and $\uparrow$ states of the bonding orbital, with an energy splitting $\Delta E_q$ determined by the external magnetic field $B$ and the hyperfine interaction between the electron and the nuclear spins of each P atoms (black arrows). The qubit electrically couples to the cavity at a resonant frequency $f_r$ via a gate defined by STM lithography.\label{scheme}}
\end{figure}	
	
	\section{THEORY}
	
    Previous theoretical analyses of qubit-cavity coupling in Si focused mostly on double quantum dot systems, using an effective Hamiltonian framework and input-output theory \cite{Hu2012, Benito2017, Burkard2020}. In DQD devices, the spin-charge hybridisation is achieved due to an inhomogeneous magnetic field that is usually created by a micromagnet deposited on the surface. The magnetic field gradient between the dots is included in the Hamiltonian as a term mixing electron spin and spatial degrees of freedom which enables spin-photon coupling.\\

	The nature of spin-charge hybridisation is different in 1P-1P devices compared to DQD, since it arises from the hyperfine interaction between the electron and the $^{31}$P nuclear spins. Other works discussing spin-charge coupling in Si:P devices include flip-flop qubit \cite{Tosi2017, Boross2018, Hetenyi2019} and 1P-2P system \cite{arxiv1P2P, Felix2021}. Those proposals consider an asymmetry in the hyperfine interaction between the left and right dot, while our work investigates a symmetric (1P-1P) case, and demonstrates that this also produces viable spin-charge coupling. Our focus here is on a device design for optimized coupling to superconducting resonator for long distance two-qubit coupling, hence we discuss its feasibility in terms of all the relevant system parameters and cavity design. Because the HF interaction involves both electron and nuclear spins, it is appropriate to include the nuclear spins in the basis and solve the problem for the joint electron-nuclear system. In Si:P devices this approach is straightforward to adopt thanks to the limited number of $^{31}$P donors to consider. 
	
    Here, we formulate the effective 1P-1P Hamiltonian in the $|D I_L I_R S\rangle$ basis, where $D$ defines an electron localized on the left or right donor ($|L\rangle$ or $|R\rangle$), $I_L$ and $I_R$ indicate the left and right nuclear spin (with a polarisation $|\Uparrow\rangle$ or $|\Downarrow\rangle$) and $S$ the electron spin ($|\uparrow\rangle$ or $|\downarrow\rangle$). We use the following 16$\times$16 Hamiltonian in the described basis:
	\begin{equation}\label{eq_H}
	H = -t_c \tau_x +\epsilon\tau_z + h\gamma_e \mathbf{B}\cdot\mathbf{S} + \sum_{j=L,R} h\gamma_P \mathbf{B}\cdot\mathbf{I}_j + H_{HF}
	\end{equation}
	where $h$ is Planck's constant, $t_c$ is the tunneling rate between the two donors, $\epsilon$ is the relative detuning energy between the left and right donor states, $\mathbf{B}=(0, 0, B)$ is an external applied magnetic field, and $\gamma_e=27.97$ GHz/T, $\gamma_P=-17.23$ MHz/T are the electron and nuclear spin gyromagnetic ratios, respectively. The Pauli matrices in the left/right donor basis are $\tau$, while $\mathbf{S}=\frac{1}{2} \{\sigma^e_x,\sigma^e_y,\sigma^e_z \}$ and $\mathbf{I}=\frac{1}{2} \{ \sigma^P_x,\sigma^P_y,\sigma^P_z \}$ are the electron and donor spin operators with $\sigma^e$ ($\sigma^P$) being the Pauli matrices in the electron (donor) basis. $H_{HF}$ is the Hamiltonian describing the HF interaction between electron and nuclear spins. It can be expressed as:
	\begin{equation}\label{eq_HHF}
	H_{HF} = 
	hA_L \mathbf{I_L}\cdot \mathbf{S}\otimes |L\rangle \langle L|
	+
	h A_R \mathbf{I_R}\cdot \mathbf{S}\otimes |R\rangle \langle R|	
	\end{equation}
	where $A_L$ ($A_R$) represents the hyperfine constant of the left (right) donor, assumed to be equal to the bulk value $A_L=A_R=117$\,MHz. The electron-nuclear spin product can be expressed as $\mathbf{I}\cdot \mathbf{S} = I_zS_z + \frac{1}{2}(I_+S_- + I_-S_+)$.\\
	
    Near zero detuning, the tunnel coupling $t_c$ mixes the $|L\rangle$ and $|R\rangle$ states to create the bonding $|-\rangle$ and antibonding $|+\rangle$ states with $|\mp\rangle = (|L\rangle\pm|R\rangle)/\sqrt{2}$. 
    We assume that the cavity field, with amplitude $\mathcal{E}_0$ and frequency $f_r$, has a non-zero polarisation component along the $x-$axis in Fig. \ref{scheme}, which is defined to be the 1P-1P axis. 
    The operator $e \mathcal{E}_0 {\bf x}$ mixes the bonding $|-\rangle$ and antibonding $|+\rangle$ states, with a charge coupling rate $g_c$ defined as $\hbar g_c \equiv \mathcal{E}_0 d_c = e \mathcal{E}_0|\langle -|  x |+\rangle|$. In the $\{ |L\rangle, |R\rangle \}$ basis we can express the coupling Hamiltonian as:  
	\begin{equation}\label{eq_Hc}
	H_c = \hbar g_c (a +  a^{\dagger})\tau_z
	\end{equation}  
	where $a$ ($a^{\dagger}$) is the annihilation (creation) operator for the microwave resonator mode. This Hamiltonian $H_c$ does not directly mix spins since it is diagonal in nuclear and electron spin subspaces, $i.e.$ $\langle D_1 I_i I_j S_a| H_c | D_2 I_l I_k S_b \rangle = \delta_{il}\delta_{jk}\delta_{ab} \langle D_1| H_c | D_2\rangle$. However, as the hyperfine  interaction is spatially-dependent (see Eq. \ref{eq_HHF}) the $H_{HF}$ induces hybridization of spin and orbital degrees of freedom in the system. We can expect this mixing to provide non-zero $g_s$ values, $i.e.$ $H_c$ evaluated between eigenstates of opposite electron spin $\hbar g_s = |\langle \Psi_i |H_c| \Psi_j \rangle|$, and thus to enable the electrically-driven spin rotations.

	The dipole moment $d_c$ and tunneling coupling $t_c$ are two important parameters to consider when discussing the spin-photon coupling $g_s$. The dipole $d_c$ determines the rate of charge-photon coupling $g_c$, which sets an upper limit for $g_s$ as the two coupling rates are proportional to each other. Tunneling is responsible for $|L\rangle$ and $|R\rangle$ state mixing, which is necessary for efficient electron movement between the dots or donors. This movement leads to periodic changes in the magnetic field for DQD or the hyperfine interaction for 1P-1P, which couples the spatial and spin degrees of freedom to enable electrically-driven spin rotations.
	We look for a transition that involves electron $|\downarrow\rangle$ and $|\uparrow\rangle$ states of the bonding orbital, $i.e.$ our qubit states (see Fig. \ref{scheme}). The cavity frequency should be matched to the energy splitting between those states -- which is on the order of the electron Zeeman splitting $\gamma_e B$. As the proximity of the antibonding orbital to the qubit levels maximizes the degree of spin-charge hybridization, we aim also to set the tunneling rate $2t_c/h$ close to the qubit energy splitting. Considering the standard cavity frequency bandwidth range of 4-12 GHz, both DQD and 1P-1P systems should be designed in a way that allows $\gamma_e B$ to be in that interval, and $2t_c/h$ to be close to that value.

    We perform atomistic calculations using the NEMO3D package \cite{nemo_ref_1, nemo_ref_2} to estimate both the tunneling rate $t_c$ and dipole moment $d_c$ that can be achieved in 1P-1P systems. In order to elucidate the challenges and feasibility of cavity-coupling to donors, we also model DQD systems using the same method as above for comparison. For both systems we use a nearest-neighbor tight-binding Hamiltonian within the $sp^3d^5s^*$ band structure, where the positions of the P donors are defined with exact lattice site precision. The donors are represented by Coulomb potentials of a single positive charge -- screened by the silicon dielectric constant $\epsilon_{Si}=11.9$ and having a cut-off value of $U_0=3.782$ eV \cite{Ahmed2009}.
	Quantum dots are represented by an external electrostatic potential of the form:
	\begin{equation}
	W(x,y,z) = \min_{i = 1,2} \, \,  (c_{xi}(x-x_i)^2+c_{yi}(y-y_i)^2)+E_z z
	\end{equation}
	where $c_{xi}$ and $c_{yi}$ are the curvatures of the $i$-th dot potential along $x$ and $y$ directions respectively and ($x_i, y_i$) is the $i$-th dot center. $E_z$ is the electric field applied from a top gate which confines the electron close to the surface in the $z$ direction. We have used symmetric dots of potential curvatures  $c_{xi} = c_{yi} = 1 \times 10^{-7}$ V/nm$^2$ and an electric field in $z$ direction of $E_z=10$ MV/m. This curvature corresponds to a dot radius of $35$ nm, as calculated from the harmonic oscillator potential, $W(x) = \frac{1}{2}m\omega ^2 x^2$, with the ground state proportional to $\exp(\frac{-x^2}{r_0^2})$ and $r_0 = \sqrt{\frac{2\hbar}{m\omega}}$. At this curvature, the harmonic oscillator energy level separation is equal to $\hbar \omega = 0.12367$ meV. This orbital energy, as well as the valley splittings found for this QD confinement, are much larger than the tunnel couplings considered in the manuscript of the order of 10 GHz $\sim$ 40 $\mu$eV. The same argument applies to donors, and therefore these higher excited states are ignored in the following.

	\begin{figure}
    	\vskip 4mm
    	\includegraphics[width=8.6cm]{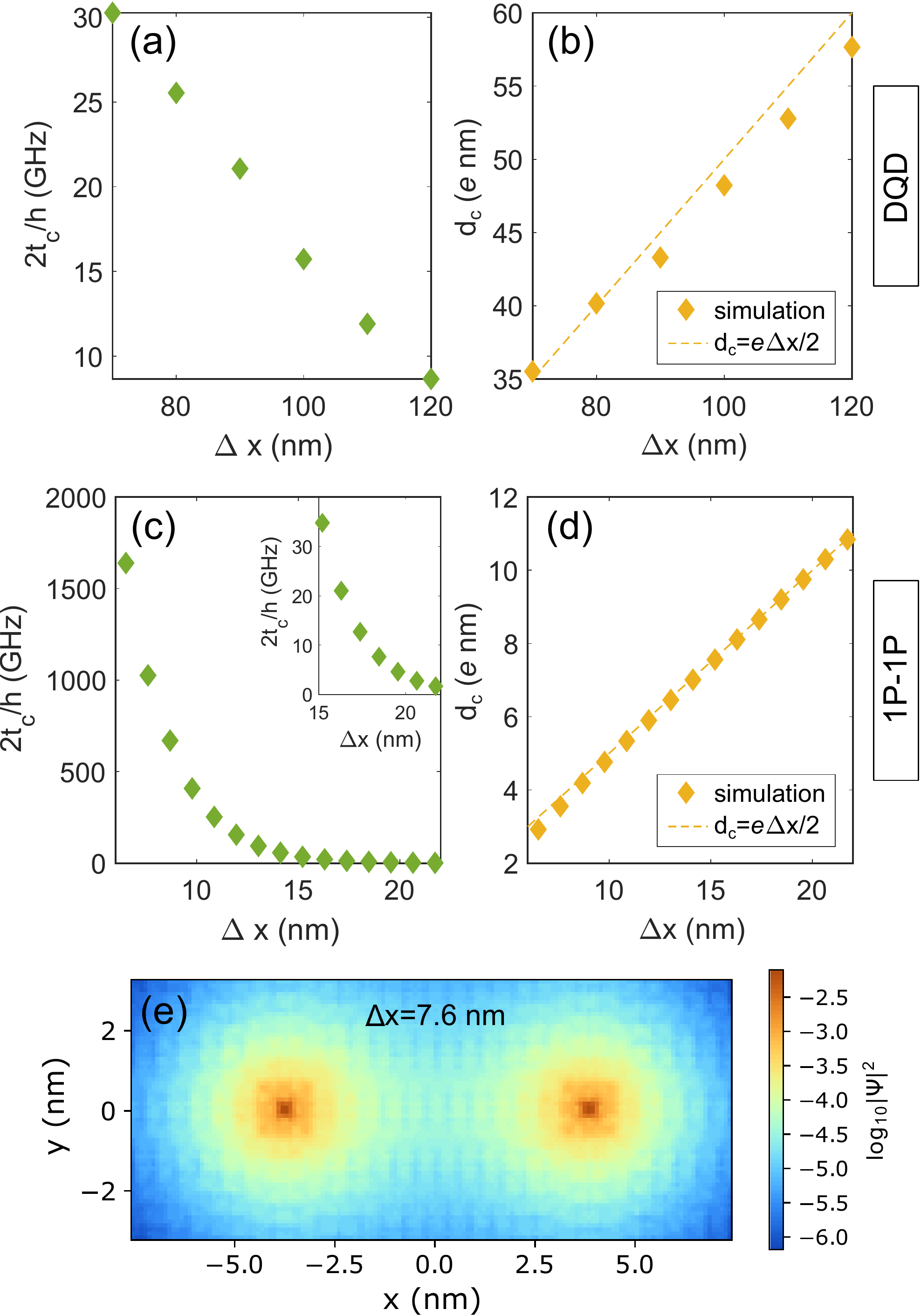}%
		\caption{ Tunneling rate $2t_c$ (a,c) and dipole moment $d_c$  (b,d) of a double quantum dot (a,b) and two donor system (c,d) as a function of dot or donor separation $\Delta x$. Dashed lines in (b,d) show relation expected for the case of ideal two-point dipole, i.e. $d_c=e\Delta x/2$. (e) Charge density of an electron localized within donors separated by 7.6 nm -- for clarity plotted as logarithm of $|\Psi|^2$. Plot shows the cross-section of the device exactly at the donors' position in $z$ direction. \label{donor_tc_d}}
	\end{figure}
	
	\section{RESULTS}
	\subsection{Tunneling and dipole moment}
	
	Using the atomistic model we calculate the tunneling energy and dipole moment for DQD (Fig. \ref{donor_tc_d}(a-b)) and 1P-1P (Fig. \ref{donor_tc_d}(c-d)) systems as a function of dot or donor separation $\Delta x$. The tunneling rate $2t_c/h$ represents the energy difference between the two lowest orbitals obtained with the tight-binding model, i.e. the bonding and antibonding orbitals. The range of $\Delta x$ has been chosen to yield tunnel rates on the order of 10 GHz, to match the typical frequency range of coplanar microwave resonators. For the DQD system, the results correspond to recent experiments \cite{Mi2018, Samkharadze2018} in which $2t_c/h$ on the order of 10 GHz has been reported for dot separation of about 100 nm.

	In the 1P-1P case, the electron wave function is significantly  more localized than for quantum dots. The electron is strongly confined to donor regions (see Fig. \ref{donor_tc_d}(e) where $|\Psi|^2$ is plotted in logarithmic scale for better visibility) so the tunneling is considerably smaller than in quantum dots for the same $\Delta x$. For instance, the tunneling of $\sim$30 GHz is obtained for dots separated by 70 nm, while the same tunneling for 1P-1P system can be achieved only if bringing donors $\sim$15 nm apart. The apparent qualitative difference between 1P-1P and DQD plots is due to the different $\Delta x$ range to wavefunction size ratio in both systems, as well as the specific choice of the dot potential profile. While in DQD the tunneling is decreasing linearly within the $\Delta x$ range we investigate (70-120 nm), for donors the drop in $t_c$ is more abrupt, demonstrating a clear exponential decay within as small $\Delta x$ range as 15 nm. To operate in a $2t_c/h$ region comparable to cavity frequency bandwidth we should focus on donor separation of about 15-20 nm.

	\begin{figure*}[t]
		\includegraphics[width=17.2cm]{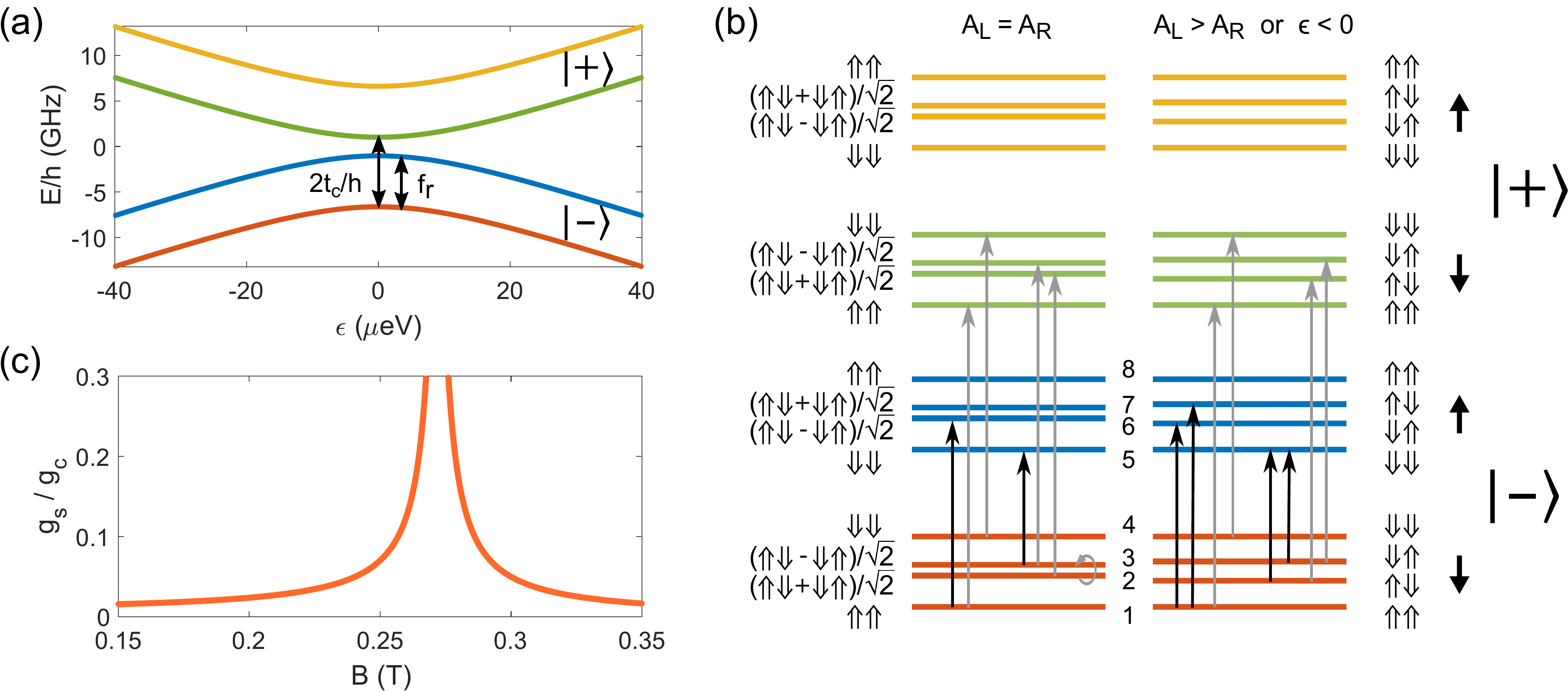}%
		\caption{(a) Energy spectra of 1P-1P system as a function of detuning, as calculated within the effective Hamiltonian approach. Each line here is nearly 4-fold degenerate due to different nuclear spins configurations. (b) Electron-nuclear energy levels for a symmetric ($A_L=A_R$) and asymmetric ($A_L>A_R$ or $\epsilon < 0$) hyperfine interaction. The color coding matches that of plot (a). Symbols $\{ \Uparrow, \Downarrow\}$, $\{\uparrow,\downarrow\}$, $\{|-\rangle, |+\rangle\}$ describe the dominant contributing state of each eigenstate. Labels 1-8 enumerate the eigenstates. Black (grey) arrows indicate electron spin flipping (conserving) transitions. (c) Spin-photon coupling $g_s$ expressed in $g_c$ units as a function of the external magnetic field $B$ for $2t_c/h = 7.64$ GHz.\label{donor_trans}}
	\end{figure*}

	The calculated dipole moments shown in Fig.~\ref{donor_tc_d}(b, d) follow approximately linear dependence as a function of the separation $\Delta x$, for both donors and DQDs. It can be noticed the simulated values are in good agreement with the classical two-point dipole $d_c{=}e\Delta x/2$ (yellow dashed line in Fig.~\ref{donor_tc_d}(b, d)), which we use in the following.

	The requirement for the tunnel coupling to closely match the cavity frequency determines the desired separation between the donors or the QDs. This consideration is unfavourable for donor systems because the separation of interest to match tunnel coupling of 10\,GHz is smaller than that of QDs -- hence resulting in a smaller charge dipole in the 1P-1P system in comparison to DQD. However this does not rule out the possibility of obtaining comparable charge-photon coupling $g_c$ by maximising the cavity field $\mathcal{E}_0$ at the donor sites. The strong localization of the electron wave function and the atomic precision of STM lithography allow the 1P-1P system to be placed in a high-field region of the cavity, as further explained in Section IV.

	We also investigate the case of an electron bound to an asymmetric donor cluster system, namely the 1P-2P configuration. 
	Because a 2P cluster presents a stronger localization potential to the electron compared to 1P, a detuning is required to bring the left and right donor levels on resonance in order to estimate a tunneling rate and dipole moment. This is done in the tight-binding model by adding a linear potential drop between the donors. We set the donors of the 2P cluster to be separated by 2 lattice constants, i.e. 1.08619 nm, along the $y$-axis. Like for the 1P-1P case, the dipole is still approximately proportional to $\Delta x/2$. Yet 1P-1P and 1P-2P systems present markedly different tunneling rates for the same cluster separation, $e.g.$ $2t_c/h$ for 1P-2P (1P-1P) is equal 38.5, 8.8, 0.4 GHz (252.7, 94.4, 12.7 GHz) for $\Delta x$ equal 10.86, 13.03, 17.38\,nm, respectively. This considerable decrease of tunneling rate for 1P-2P with respect to 1P-1P can be intuitively understood as a result of a deeper overall donor potential and therefore stronger electron localization. This is undesirable for qubit-cavity coupling, since setting $2t_c$ resonant with a cavity would require bringing the 1P-2P cluster closer together, leading to a smaller dipole moment. However, that problem can be solved by loading the clusters with more electrons, $e.g.$ 3 electrons for 1P-2P system~\cite{Felix2021}.

	\subsection{Hyperfine-mediated spin-photon coupling}

We use the effective Hamiltonian defined in Eq. (\ref{eq_H}) to calculate the energy spectra of a 1P-1P system. We show in Fig.~\ref{donor_trans}(a) the calculated energy levels of the system as a function of the detuning $\epsilon$ for two donors separated by $\Delta x = 18.47$\,nm, in an applied external magnetic field $B=0.2$\,T. The tunneling rate for this separation, as shown in Fig. \ref{donor_tc_d}(c), is equal to $2t_c/h = 7.64$\,GHz. Each of the four lines in Fig.~\ref{donor_trans}(a) is itself a group of four distinct eigenstates, illustrated in Fig.~\ref{donor_trans}(b). The left side of Fig.~\ref{donor_trans}(b) shows the spectrum for a symmetric ($A_L=A_R$) hyperfine interaction, and the right side for an asymmetric ($A_L>A_R$ or equivalently $\epsilon<0$) interaction. The four eigenstates in each manifold differ only by nuclear spin configuration, and their order is determined by the HF interaction, as for $B$=0.2 T it dominates over the nuclear Zeeman interaction.

For ease of notation, the eigenstates in Fig. \ref{donor_trans}(b) are only labelled by their dominant contributing state, and the labels do not show other basis states that have a finite weighting in the eigenstates. In fact, due to the off-diagonal terms in the hyperfine Hamiltonian $H_{HF}$, the states labelled in Fig. \ref{donor_trans}(b) are also composed of admixtures of other spin and orbital basis states. The nature of those admixtures can be predicted from the dominant contributing state and the form of $H_{HF}$, as HF causes electron-nuclear spin flip-flops but does not change the total spin or the overall parity of the states. While the electric field of the resonator interacts with the 1P-1P via the charge dipole matrix, which couples the bonding and anti-bonding orbital states, and thus changes the overall parity of the state.

We further estimate the spin-photon coupling $\hbar g_s$ by evaluating $H_c$ between the $H$ eigenstates at detuning $\epsilon=0$. Non-zero matrix elements $\langle\psi_i|H_c|\psi_j\rangle$ indicate that $i$-th and $j$-th states are coupled by a perturbation from the cavity electric field. That determines which transitions are possible, provided the cavity frequency $f_r$ is resonant with the given level's energy splitting.
	 
We can detail the composition of the ground state $\psi_1$, and of the 6-th eigenstate $\psi_6$, to understand how the admixtures of different spin and orbital states can result in electrically available transitions: 
	\begin{equation}
	\begin{aligned}
	    |\psi_1\rangle = {} & -0.7071(|L\rangle + |R\rangle) |\Uparrow\Uparrow\downarrow\rangle \\ 
	    & + 0.0035(|L\rangle + |R\rangle) (|\Uparrow\Downarrow\rangle +|\Downarrow\Uparrow\rangle)|\uparrow\rangle \\
	    & - 0.0015(|L\rangle - |R\rangle) (|\Uparrow\Downarrow\rangle-|\Downarrow\Uparrow\rangle)|\uparrow\rangle, \\ 
	    |\psi_6\rangle = {} & -0.4999(|L\rangle + |R\rangle) (|\Uparrow\Downarrow\rangle-|\Downarrow\Uparrow\rangle)|\uparrow\rangle \\
	    & + 0.002(|L\rangle - |R\rangle) (|\Uparrow\Downarrow\rangle+|\Downarrow\Uparrow\rangle)|\uparrow\rangle \\
	    & -0.0146(|L\rangle - |R\rangle) |\Uparrow\Uparrow\downarrow\rangle,
	\end{aligned}
	\end{equation}
																						 
    In the above, non-zero transition elements driven by $H_c$ appear between same-spin, opposite-orbital contributions from both eigenstates, $i.e.$ between the first term of $\psi_1$ and the last term of $\psi_6$ etc. Note that this can be understood as the electric drive exciting the orbital basis from $|-\rangle$ to $|+\rangle$.
    Indeed, taking all the terms into account, we obtain a finite expected value $\hbar g_s^{1,6} = |\langle\psi_1|H_c|\psi_6\rangle| \approx 0.023\,\hbar g_c$. The same approach can be applied to identify all the other possible electron spin-flipping transitions, which are represented by the black arrows in Fig. \ref{donor_trans}(b). Note that these electrical transitions in tunnel-coupled donor systems can in principle also be used for EDSR, and parallel approaches have investigated the advantages of asymmetric clusters and multi-electron systems in this context \cite{arxiv1P2P, Felix2021}. We ignore transitions to the antibonding electron-spin up states because of their significant energy separation with respect to the resonances considered in this paper. The transition matrix elements between states of the same spin but different spatial symmetry are equal to approximately $\hbar g_c$. As mentioned above, the transitions which include an electron spin rotation always involve a nuclear spin transition as well - with the total spin of the joint electron-nuclear system being conserved. The configurations where the nuclear spins are polarised in the same direction, $e.g.$ $\Downarrow\Downarrow\downarrow$, do not result in an available cavity-driven electron-spin transition because there is no possible flip-flop transition, and the system is in the equivalent of a blockade regime.

    A distinction must be made between the symmetric and asymmetric hyperfine cases. For the symmetric case, the configurations $\Uparrow\Downarrow$ and $\Downarrow\Uparrow$ form a nuclear spin singlet and triplet. This results in the state $\Psi_1$ (with dominant contribution of $\Uparrow\Uparrow$) to only couple to the nuclear singlet of opposite electron spatial symmetry and spin $\Psi_6$ (of dominant contribution of $\frac{1}{\sqrt{2}}(\Uparrow\Downarrow-\Downarrow\Uparrow)$), but not to the nuclear triplet $\Psi_7$.
	In the case of an asymmetric hyperfine interaction between the left and the right donors, the nuclear singlet-triplet symetry is broken as one of the configurations $\Uparrow\Downarrow$ or $\Downarrow\Uparrow$ is energetically favoured, as depicted on the right-hand side of Fig. \ref{donor_trans}(b). Thus for asymmetric hyperfine there are four possible electrical transitions while for the symmetric case only two are allowed.

    The splitting between $\Uparrow\Uparrow$ and $\Downarrow\Downarrow$ nuclear configuration, within the $|-\rangle$ or the $|+\rangle$ orbital, is a sum of hyperfine and nuclear Zeeman energies $(A_L+A_R)/4 \pm 2\gamma_PB$. As at the point of anticrossing the eigenstates contain some admixtures of different orbital and spin states, the actual energy splittings slightly deviate from the values mentioned above. For instance, the $\Uparrow\Uparrow$ and $\Downarrow\Downarrow$ levels in the lowest quadruplet differ by 65.8 MHz while for non-mixed states we would expect a splitting of 65.4 MHz. Also, the degeneracy between the nuclear spin singlet and triplet is lifted by about $0.1$ MHz. In case of any asymmetry in the hyperfine interaction between the left and the right cluster, one of the configurations $\Uparrow\Downarrow$ or $\Downarrow\Uparrow$ is energetically favoured and the nuclear-electron levels are further split by about $|A_L-A_R|/4$.
	
    Since the antibonding $|+\rangle$ orbital plays a mediating role in the considered transition, the $g_s$ coupling will depend on how strongly the orbitals are mixed and, consequently, on the relative values between $2t_c$ and the Zeeman splitting. We calculate $g_s/g_c$ as a function of the magnetic field $B$ while keeping $t_c$ constant -- see Fig. \ref{donor_trans}(c). We can see that the spin-photon coupling increases when the electron spin Zeeman splitting approaches $2t_c$ at about 0.272 T, that is the degeneracy point between the $|-\rangle\uparrow$ and $|+\rangle\downarrow$ states. Although $g_s$ is highest there as it equals $g_c$, this point is also a hot-spot where the spin qubit is most susceptible to charge decoherence --  similarly to DQD systems.

 	\subsection{Electrically induced spin-orbit interaction}
	
    Now we consider the effect of spin-orbit interaction on the donor system. As recently reported~\cite{weber2018spin}, a pronounced spin-orbit coupling  arises in donor qubit devices in Si when placing P donors in a strong static electric field. It has been shown in analysis of single donor electron spin-relaxation processes~\cite{weber2018spin} that this effect, referred here as EISO, can dominate over the Rashba and bulk silicon crystal SO interactions. In following we focus only on EISO effect. In a 1P-1P system, the EISO mechanism combined with an electric field difference between the two donor sites becomes equivalent to a magnetic field gradient used in DQDs to electrically induce spin-flip transitions.
    We include the following additional term $H_{EISO}$ in the effective Hamiltonian $H$ of eq. (\ref{eq_H}):
	
	\begin{figure}
		\includegraphics[width=8.6cm]{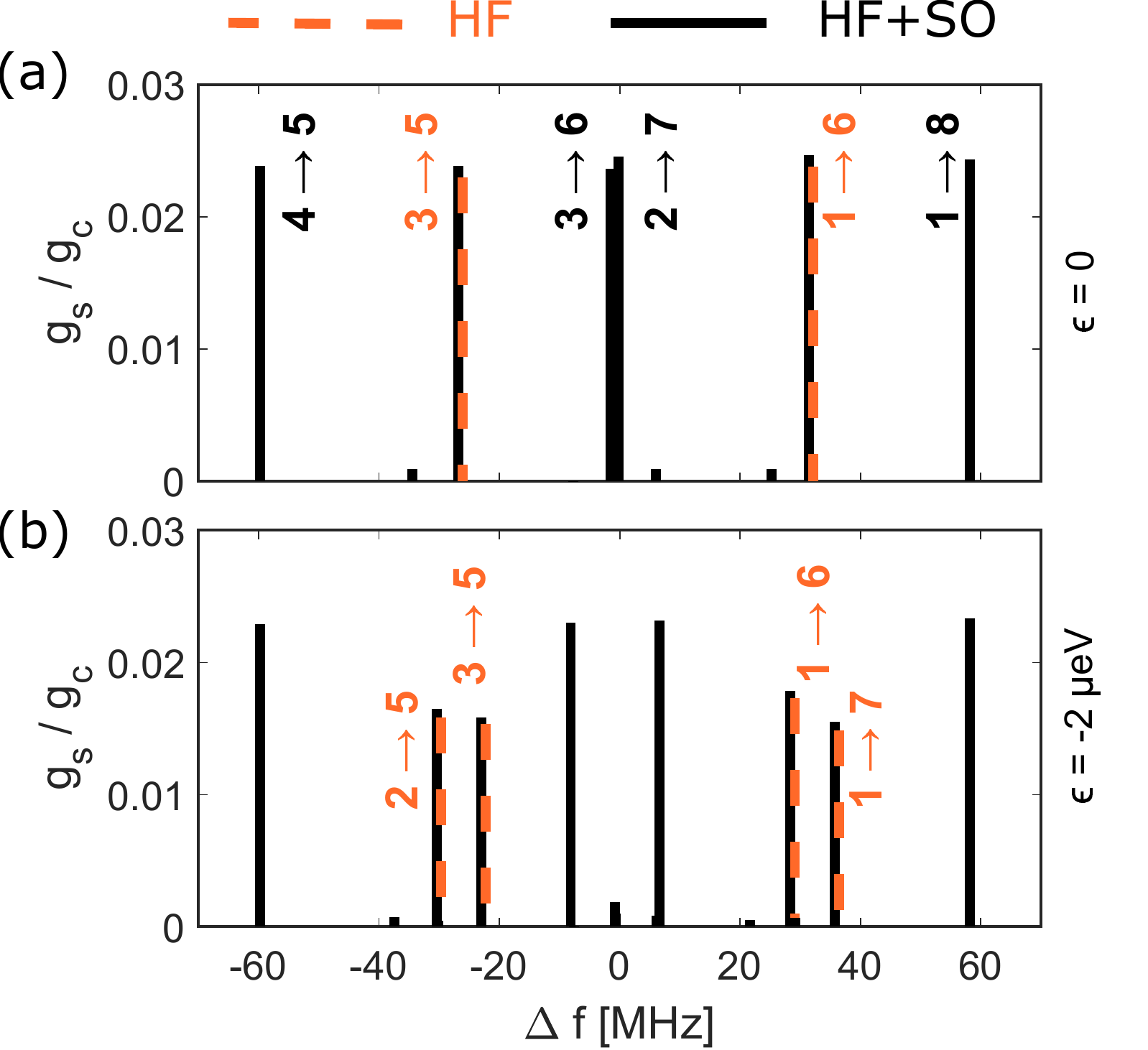}%
		\caption{Spin-photon coupling $g_s$ expressed in $g_c$ units for all possible transitions between the first two quadruplets (states 1-4 and 5-8). $g_s$ is represented by a peak at each transition's resonant frequency, with the $x$ axis representing the frequency shifted by the electron Zeeman energy $\Delta f = f_r - \gamma_e B$. The black solid (orange dashed) line shows results for both the hyperfine and the  spin-orbit (hyperfine only) interaction included in the Hamiltonian. Plots (a) and (b) correspond to the cases of $\epsilon = 0$ and $\epsilon = -2\, \mu eV$, respectively. These results were calculated for a magnetic field $B_z=0.2$ T and external electric fields $E_y^L=15$ MV/m, $E_y^R=-15$ MV/m. \label{donor_so_peaks}}
	\end{figure}

	\begin{equation}
	H_{EISO}^e=
	\begin{pmatrix}
	E_yB_z + E_zB_y \\
	E_zB_x + E_xB_z \\
	E_xB_y + E_yB_x
	\end{pmatrix}^T
	\cdot \mathbf{C} \cdot \mathbf{\sigma}
	= E_yB_zC\sigma_x
	\end{equation}	
	where $E_x$, $E_y$, $E_z$ are the external electric field vector components and $\mathbf{C}$ is a tensor representing the strength of the spin-orbit coupling. We can notice that only an electric field transverse to the applied magnetic field can produce non-zero elements in the EISO Hamiltonian. For simplicity, we consider that the electric field is applied along the $y$-axis, $i.e.$ transverse to the $z$-axis external magnetic field. Based on spin-relaxation anisotropy measurements, the $C$ coefficient has been estimated to about $6\times10^{-14} \ e$m/T \cite{weber2018spin}. Because a difference in the SO strength is required between the left and right donor to obtain a finite spin-photon coupling, we assume different values of electric field $E_y^L$ and $E_y^R$ at the left and right donor sites. This spin-orbit Hamiltonian then reads:
	\begin{equation}
	H_{EISO}=
	\begin{pmatrix}
	E_y^LB_zC & 0 \\
	0 & E_y^RB_zC
	\end{pmatrix}
	\otimes \sigma_x^e
	\end{equation}
	where the first matrix is in the $\{ |L\rangle, |R\rangle \}$ basis. This SO mechanism has no influence on the nuclear spins, and the Hamiltonian is diagonal in the nuclear spin basis. Therefore, $H_{EISO}$ enables electrically driven electron spin rotations whilst preserving nuclear polarisation, $e.g.$ $|\Uparrow\Uparrow\downarrow\rangle \rightarrow |\Uparrow\Uparrow\uparrow\rangle$, which is not possible via the HF interaction. Focusing on this particular transition, at a magnetic field $B_z=0.2$ T, we obtain $g_s/g_c\approx 1\%$ when the electric field difference between the dots is set to 12 MV/m ($E_y^L=6$ MV/m, $E_y^R=-6$ MV/m), similar to those obtained with the HF mechanism. While the absolute values of these electric fields are reasonable for donor qubit devices, the opposite signs imply a steep field gradient that can be challenging to engineer in practice. This EISO coupling, however, presents an advantage over the HF mechanism in not requiring spin flip-flops for electron spin rotations, and therefore enable spin-photon coupling regardless of the initial nuclear spin state.

	In Fig. \ref{donor_so_peaks}, $g_s/g_c$ is shown for the possible transitions between pairs of eigenstates from Fig. \ref{donor_trans} (b).
	The calculations were done using  $B_z=0.2$ T, $E_y^L=15$ MV/m, $E_y^R=-15$ MV/m, with these electric field values chosen so that HF and EISO $g_s$ would be of similar magnitude. For clarity, the resonances are plotted as a function of the frequency shifted by the electron Zeeman splitting $\Delta f = f_r - \gamma_e B$. The orange dashed lines represent the available transitions obtained when only the HF interaction is considered, while the black lines represent the transitions when both HF and EISO mechanisms are included. Both the zero and non-zero detuning cases are considered, and the numbers above each line indicate the eigenstates between which the transitions occur.
	The transitions labelled 1$\rightarrow$8, 2$\rightarrow$7, 3$\rightarrow$6 and 4$\rightarrow$5 are enabled by EISO only, while
	the small visible peaks of $g_s/g_c$ on the order of 0.001 correspond to transitions for which both HF and EISO interaction are needed. These results clearly show that EISO offers additional means of coupling spins to microwave cavity, with the benefits of being insensitive to the nuclear polarisation and relatively independent of HF-related $g_s$ coupling.

	\subsection{Qubit operation and decoherence}
	
	The HF and EISO interactions enable different transitions in Fig. \ref{donor_so_peaks}, at distinct resonant frequencies, and each of these frequencies can be treated as a possible working point for electron spin qubit manipulation.  The separation between HF- and EISO-enabled resonances is on the order of $(A_L+A_R)/8$, or tens of MHz, and it should be possible to distinguish the peaks originating from those two mechanisms with the frequency resolution offered by current cavity architectures.

	\begin{figure}
	\includegraphics[width=8.6cm]{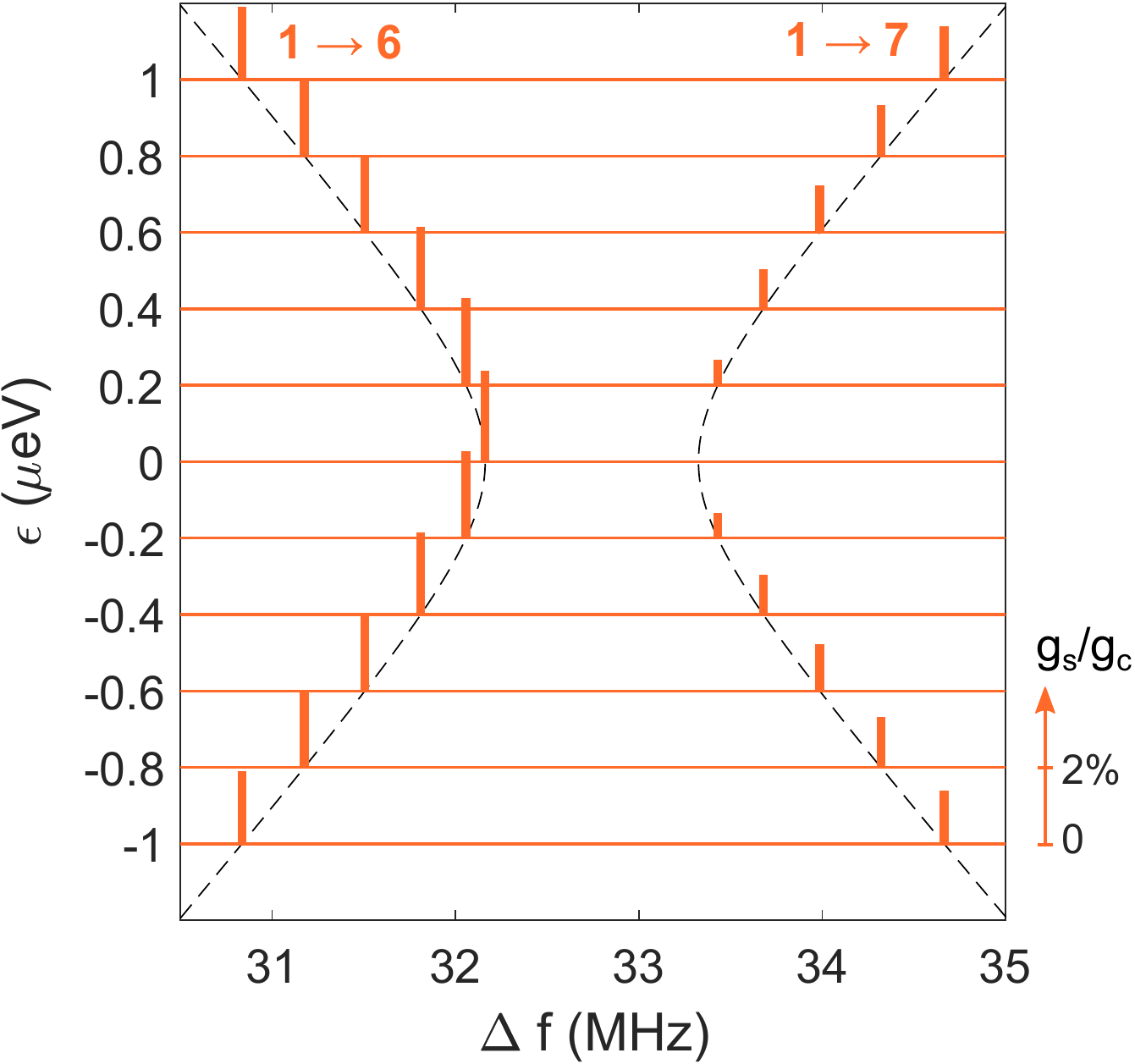}%
	\caption{Effect of detuning on the frequencies $\Delta f = f_r - \gamma_e B$ of HF-driven resonances $1\rightarrow6$ and $1\rightarrow7$. Height of the peaks represent $g_s/g_c$ values corresponding to each resonance. \label{hf_detuning}}
	\end{figure}
	
	Assuming all nuclear spin orientations have approximately the same probability (reasonable for temperatures on the order of 100\,mK), the initial qubit state will appear in each of the four possible quadruplet eigenstates in the lowest branch of Fig. \ref{donor_trans}(b) equally often. Similarly to ESR experiments \cite{Wang2016}, at any given time we would observe only resonances corresponding to the current nuclear spin state. Additionally, as explained previously, not all initial qubit states are sensitive to electric field when solely the HF interaction is present in the system. Only including the EISO mechanism yields electrical transitions for any initial nuclear spin state. For example, the initial spin state $|\Downarrow\Downarrow\downarrow \rangle$ is insensitive to electric field for HF alone, but has an allowed transition due to the EISO mechanism, see peak $4 \rightarrow 5$ in Fig. \ref{donor_so_peaks}(a). The problem of non-responsive initial states for the HF-only case could be overcome with nuclear spin polarisation methods like NMR \cite{Pla2013} or DNP \cite{Abragam1978, Simmons2011}. Control over initial nuclear spin configuration, provided by these methods, would allow a single chosen resonance in Fig \ref{donor_so_peaks} to be deterministically addressed.
	
    Single donor qubits can achieve very long coherence and relaxations times due to large excited state energies and the absence of spurious nuclear spins in isotopically purified silicon \cite{Muhonen2014,Watson2017}. The presence of a spin-orbit mechanism, necessary to enable electrical qubit manipulation, also exposes the qubit to charge noise which can become the main source of decoherence \cite{Yoneda2018}, and make the strong spin-photon coupling regime challenging to achieve \cite{Petersson2012}. However, donor qubits can be embedded in a fully epitaxial structure while quantum dots are by definition pinned to an interface, usually rich in two-level fluctuators. Hence, noise figures as low as $0.094 \ \mu$eV$/\sqrt{Hz}$ (at 1 Hz) have been evidenced for a 2P-3P system \cite{Kranz2020}, against a range of $0.3 - 2 \ \mu$eV$/\sqrt{Hz}$ for electrostatically-defined quantum dots \cite{SchreiberNJP20}. Together with the smaller dipole for donor systems at fixed tunneling frequency, we can therefore expect a smaller or at least similar spin decoherence rate for a 1P-1P system strongly coupled to a microwave resonator compared to the 1-2 MHz range measured for DQDs \cite{Samkharadze2018, Mi2018}, which should facilitate the achievement of the strong coupling regime.

    Charge noise generates uncontrollable fluctuations in detuning, which shifts the qubit frequency and introduces phase decoherence. For the 1P-1P system we consider, detuning also causes asymmetry in the HF interaction, shifting the energies of some nuclear spin states. This noise-induced asymmetry thus results in additional electron spin-flipping transitions (see Fig. \ref{donor_trans}b with $1 \rightarrow 6$ transition available for symmetric and both $1 \rightarrow 6$ and $1 \rightarrow 7$ for asymmetric case). Defining the $1 \rightarrow 6$ transition as the qubit levels, the frequency shift and strength of the nearby $1 \rightarrow 7$ transition, as a function of detuning, are illustrated in Fig. \ref{hf_detuning}. The height of the bars represents the $g_s/g_c$ value for each resonance at a particular detuning. This electron spin transition outside the qubit subspace represents another potential decoherence mechanism due to charge noise.
    
	\begin{figure}[t!]
	\vskip 1mm
	\includegraphics[width=8.6cm]{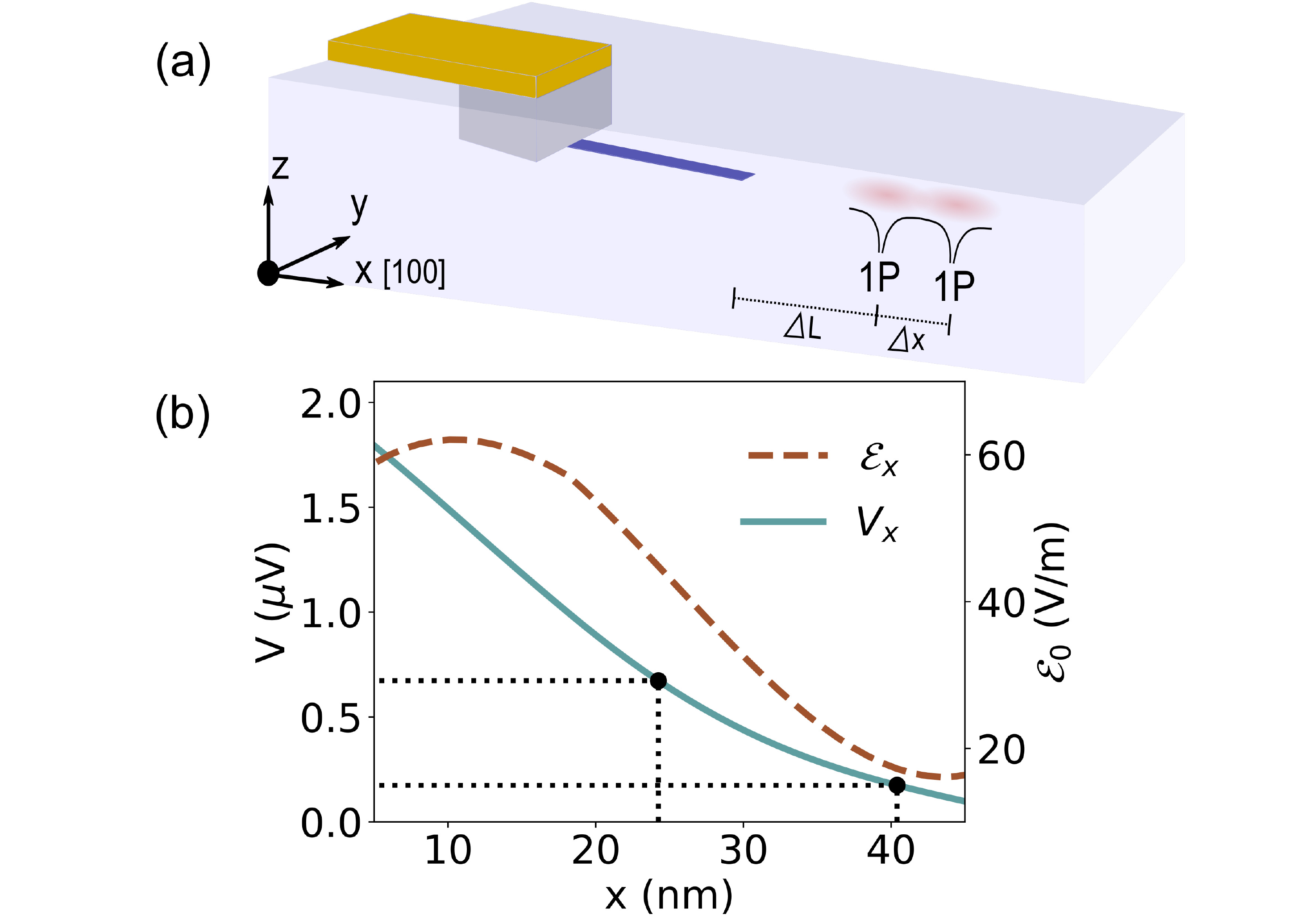}%
	\caption{ (a) An illustration of one end of a $\lambda$/2 NbTiN microwave resonator, terminating in a phosphorous $\delta$-layer lead at the qubit plane, 50 nm below the silicon surface. The 1P-1P system is separated by approximately 15-20 nm, and  placed $\Delta L > 20$ nm from the phosphorous $\delta$-layer lead, using the nanometer precision of scanning-probe lithography. These two approximate distances are chosen to ensure that the 1P-1P tunneling frequency is in the 4-12 GHz range, and the tunneling frequency from the lead to the closest donor is far slower than any other time scale in the system.
		(b) Electric field and voltage from a single photon in the resonator mode, for a configuration similar to that illustrated in (a), simulated using Ansys HFSS. The voltage values are given at the two donor positions, and show a voltage detuning of approximately $\Delta V \sim 0.5 \mu$V.
		\label{resonator_leads}}
	\end{figure}
	
	\section{Resonator simulations}

To enhance electric dipole coupling between a microwave resonator and a 1P-1P system, it is proposed to contact the ends of a resonator at the silicon surface to short phosphorous-doped leads at the qubit layer \citep{Ruess2004}, 50 nm beneath the surface, using vias as illustrated in Fig. \ref{resonator_leads}(a). In this way, the resonator anti-node can be brought to a distance slightly greater than 20 nm from one of the donors, to increase the differential lever arm.

A $\lambda/2$ superconducting microwave resonator in a notch configuration, similar in design to structures described in Ref.~\citep{VandersypenPhysRevAppl19}, was simulated using the Ansys HFSS package. The ends of the resonator were modelled to contact phosphorous $\delta$-doped leads at the qubit layer, which have a resistivity of $R_{xx} = 350 \ \Omega / \msquare$ at 50 mK \citep{Ruess2004}. The simulated $S21$ parameter was fitted using the Python package described in Ref.~\citep{Probst15}, and a loaded quality factor $Q_l \sim 5 \times 10^3$ was extracted from the fit, comparable to quality factors cited in Refs.~\citep{Mi2018, Samkharadze2018}. Simulating an intra-cavity field at the single-photon level, the resulting electric field gradient in the silicon substrate is plotted along the 1P-1P axis in Fig. \ref{resonator_leads}(b). Here, we take donor separation $\Delta x = 15-20$ nm as the correct order of magnitude ($\Delta x = 18.47$ nm has been used in spin-photon coupling simulations). We also assume a distance $\Delta L$ slightly more than 20 nm from the resonator's delta-layer termination to the nearest donor, to limit tunnel coupling between the donor and resonator lead. With these parameters and simulation of the electric field from the cavity, a donor detuning of $\Delta V \sim 0.5 \ \mu V$ is estimated in Fig. \ref{resonator_leads}(b).
The engineered detuning of the 1P-1P from the high-impedance resonator can be compared to the detuning due to charge noise $\sigma_{\epsilon} = 0.17 \ \mu eV$ (1 s) measured for double donors in epitaxial silicon \citep{Kranz2020}. The fact that $e \Delta V/\sigma_{\epsilon} > 1$ suggests that dipole coupling to the resonator would be stronger than the effect of charge noise on the qubit.

In previous work \citep{VandersypenPhysRevAppl19}, a potential difference of about 20 $\mu$V was inferred between the ends of their $\lambda/2$ resonator. A similar resonator design was used to demonstrate strong spin-photon coupling with electrostatically-defined DQDs separated by about 100 nm \citep{Samkharadze2018}. It should be noted that the quantum well providing confinement along the $z-$axis was about 50 nm beneath the silicon surface, and it can therefore be assumed that the DQDs are actually detuned by less than the maximum $20 \ \mu$V. It is also important to note that, in contrast to previous work \citep{Samkharadze2018,VandersypenPhysRevAppl19}, our Ansys HFSS simulations do not take into account the kinetic inductance of the thin superconducting film, which would further enhance the impedance of the cavity and therefore the resonant detuning at the qubit location.

The 1P-1P system described here will have a smaller electric dipole moment compared to DQDs, due to donor separation that is approximately one order of magnitude smaller for a similar tunneling rate. On the other hand, the resonator lead can be closer to donors than for DQDs, and the fact that the donors are in a region of higher field partly counter-balances the smaller electric dipole moment. 

To reach the strong-coupling regime, the spin-photon coupling rate must be greater than both the cavity linewidth and the qubit decoherence rate. Simulations suggest that a charge-photon coupling rate $g_c = e \Delta V /2 h \sim 100$ MHz and $g_s/g_c = 0.02$ is realistic in the proposed device, to give a spin-photon coupling rate of 2 MHz. A resonator with fundamental frequency of 7.64 GHz should have a loaded quality factor close to $10^4$ for a linewidth below 1 MHz, which is realistic with careful resonator design and fabrication.

	\section{CONCLUSIONS}

    We have established a complete platform to estimate the spin-photon couplings that can realistically be achieved in donor systems in silicon. Importantly, we use this framework to predict that strong spin-photon coupling is achievable in realistic donor device. We start from tight-binding wavefunction simulations to estimate the tunnel couplings and associated charge dipole couplings as a function of the inter-donor distance. Using an effective Hamiltonian approach that includes the hyperfine and EISO interactions for spin-orbit coupling, we show that electrically-driven transitions exist between eigenstates. Our device proposal includes a $\lambda/2$ microwave resonator whose ends contact phosphorus $\delta$-layer leads buried at the qubit layer, to allow the coupled donor system to be positioned in a region of high electric field gradient and thus enhance charge-photon coupling. Our simulations show that a spin-photon coupling rate of the order of 1 MHz, similar to what has been observed in DQD systems, are achievable in donor-based devices without an additional micro-magnet fabricated close to the qubit. The simplified fabrication combined with excellent coherence properties are key advantages of using donors to build scalable and electrically addressable spin qubits.

	\section*{ACKNOWLEDGEMENTS}
	This research was conducted by the Australian Research Council Centre of Excellence for Quantum Computation and Communication Technology (CE170100012), the US Army Research Office under contract number W911NF-17-1-0202 and Silicon Quantum Computing Pty Ltd.
	
	The research was undertaken with the assistance of resources and services from the National Computational Infrastructure (NCI) under an NCMAS 2020 allocation, supported by the Australian Government, and of the computational cluster Katana supported by Research Technology Services at UNSW Sydney.

	\bibliography{cavity_biblio}

\begin{thebibliography}{59}%
\makeatletter
\providecommand \@ifxundefined [1]{%
 \@ifx{#1\undefined}
}%
\providecommand \@ifnum [1]{%
 \ifnum #1\expandafter \@firstoftwo
 \else \expandafter \@secondoftwo
 \fi
}%
\providecommand \@ifx [1]{%
 \ifx #1\expandafter \@firstoftwo
 \else \expandafter \@secondoftwo
 \fi
}%
\providecommand \natexlab [1]{#1}%
\providecommand \enquote  [1]{``#1''}%
\providecommand \bibnamefont  [1]{#1}%
\providecommand \bibfnamefont [1]{#1}%
\providecommand \citenamefont [1]{#1}%
\providecommand \href@noop [0]{\@secondoftwo}%
\providecommand \href [0]{\begingroup \@sanitize@url \@href}%
\providecommand \@href[1]{\@@startlink{#1}\@@href}%
\providecommand \@@href[1]{\endgroup#1\@@endlink}%
\providecommand \@sanitize@url [0]{\catcode `\\12\catcode `\$12\catcode
  `\&12\catcode `\#12\catcode `\^12\catcode `\_12\catcode `\%12\relax}%
\providecommand \@@startlink[1]{}%
\providecommand \@@endlink[0]{}%
\providecommand \url  [0]{\begingroup\@sanitize@url \@url }%
\providecommand \@url [1]{\endgroup\@href {#1}{\urlprefix }}%
\providecommand \urlprefix  [0]{URL }%
\providecommand \Eprint [0]{\href }%
\providecommand \doibase [0]{https://doi.org/}%
\providecommand \selectlanguage [0]{\@gobble}%
\providecommand \bibinfo  [0]{\@secondoftwo}%
\providecommand \bibfield  [0]{\@secondoftwo}%
\providecommand \translation [1]{[#1]}%
\providecommand \BibitemOpen [0]{}%
\providecommand \bibitemStop [0]{}%
\providecommand \bibitemNoStop [0]{.\EOS\space}%
\providecommand \EOS [0]{\spacefactor3000\relax}%
\providecommand \BibitemShut  [1]{\csname bibitem#1\endcsname}%
\let\auto@bib@innerbib\@empty
\bibitem [{\citenamefont {Fowler}\ \emph {et~al.}(2012)\citenamefont {Fowler},
  \citenamefont {Mariantoni}, \citenamefont {Martinis},\ and\ \citenamefont
  {Cleland}}]{Fowler2012}%
  \BibitemOpen
  \bibfield  {author} {\bibinfo {author} {\bibfnamefont {A.~G.}\ \bibnamefont
  {Fowler}}, \bibinfo {author} {\bibfnamefont {M.}~\bibnamefont {Mariantoni}},
  \bibinfo {author} {\bibfnamefont {J.~M.}\ \bibnamefont {Martinis}},\ and\
  \bibinfo {author} {\bibfnamefont {A.~N.}\ \bibnamefont {Cleland}},\
  }\bibfield  {title} {\bibinfo {title} {Surface codes: Towards practical
  large-scale quantum computation},\ }\href
  {https://doi.org/10.1103/PhysRevA.86.032324} {\bibfield  {journal} {\bibinfo
  {journal} {Phys. Rev. A}\ }\textbf {\bibinfo {volume} {86}},\ \bibinfo
  {pages} {032324} (\bibinfo {year} {2012})}\BibitemShut {NoStop}%
\bibitem [{\citenamefont {Hill}\ \emph {et~al.}(2015)\citenamefont {Hill},
  \citenamefont {Peretz}, \citenamefont {Hile}, \citenamefont {House},
  \citenamefont {Fuechsle}, \citenamefont {Rogge}, \citenamefont {Simmons},\
  and\ \citenamefont {Hollenberg}}]{Hill2015}%
  \BibitemOpen
  \bibfield  {author} {\bibinfo {author} {\bibfnamefont {C.~D.}\ \bibnamefont
  {Hill}}, \bibinfo {author} {\bibfnamefont {E.}~\bibnamefont {Peretz}},
  \bibinfo {author} {\bibfnamefont {S.~J.}\ \bibnamefont {Hile}}, \bibinfo
  {author} {\bibfnamefont {M.~G.}\ \bibnamefont {House}}, \bibinfo {author}
  {\bibfnamefont {M.}~\bibnamefont {Fuechsle}}, \bibinfo {author}
  {\bibfnamefont {S.}~\bibnamefont {Rogge}}, \bibinfo {author} {\bibfnamefont
  {M.~Y.}\ \bibnamefont {Simmons}},\ and\ \bibinfo {author} {\bibfnamefont
  {L.~C.~L.}\ \bibnamefont {Hollenberg}},\ }\bibfield  {title} {\bibinfo
  {title} {A surface code quantum computer in silicon},\ }\href
  {https://advances.sciencemag.org/content/1/9/e1500707} {\bibfield  {journal}
  {\bibinfo  {journal} {Science Advances}\ }\textbf {\bibinfo {volume} {1}},\
  \bibinfo {pages} {e1500707} (\bibinfo {year} {2015})}\BibitemShut {NoStop}%
\bibitem [{\citenamefont {O'Gorman}\ \emph {et~al.}(2016)\citenamefont
  {O'Gorman}, \citenamefont {Nickerson}, \citenamefont {Ross}, \citenamefont
  {Morton},\ and\ \citenamefont {Benjamin}}]{OGorman2016}%
  \BibitemOpen
  \bibfield  {author} {\bibinfo {author} {\bibfnamefont {J.}~\bibnamefont
  {O'Gorman}}, \bibinfo {author} {\bibfnamefont {N.~H.}\ \bibnamefont
  {Nickerson}}, \bibinfo {author} {\bibfnamefont {P.}~\bibnamefont {Ross}},
  \bibinfo {author} {\bibfnamefont {J.~J.~L.}\ \bibnamefont {Morton}},\ and\
  \bibinfo {author} {\bibfnamefont {S.~C.}\ \bibnamefont {Benjamin}},\
  }\bibfield  {title} {\bibinfo {title} {A silicon-based surface code quantum
  computer},\ }\href {https://doi.org/10.1038/npjqi.2015.19} {\bibfield
  {journal} {\bibinfo  {journal} {npj Quantum Information}\ }\textbf {\bibinfo
  {volume} {2}},\ \bibinfo {pages} {15019} (\bibinfo {year}
  {2016})}\BibitemShut {NoStop}%
\bibitem [{\citenamefont {Pica}\ \emph {et~al.}(2016)\citenamefont {Pica},
  \citenamefont {Lovett}, \citenamefont {Bhatt}, \citenamefont {Schenkel},\
  and\ \citenamefont {Lyon}}]{Pica2016}%
  \BibitemOpen
  \bibfield  {author} {\bibinfo {author} {\bibfnamefont {G.}~\bibnamefont
  {Pica}}, \bibinfo {author} {\bibfnamefont {B.~W.}\ \bibnamefont {Lovett}},
  \bibinfo {author} {\bibfnamefont {R.~N.}\ \bibnamefont {Bhatt}}, \bibinfo
  {author} {\bibfnamefont {T.}~\bibnamefont {Schenkel}},\ and\ \bibinfo
  {author} {\bibfnamefont {S.~A.}\ \bibnamefont {Lyon}},\ }\bibfield  {title}
  {\bibinfo {title} {Surface code architecture for donors and dots in silicon
  with imprecise and nonuniform qubit couplings},\ }\href
  {https://doi.org/10.1103/PhysRevB.93.035306} {\bibfield  {journal} {\bibinfo
  {journal} {Phys. Rev. B}\ }\textbf {\bibinfo {volume} {93}},\ \bibinfo
  {pages} {035306} (\bibinfo {year} {2016})}\BibitemShut {NoStop}%
\bibitem [{\citenamefont {Veldhorst}\ \emph {et~al.}(2017)\citenamefont
  {Veldhorst}, \citenamefont {Eenink}, \citenamefont {Yang},\ and\
  \citenamefont {Dzurak}}]{Veldhorst2017}%
  \BibitemOpen
  \bibfield  {author} {\bibinfo {author} {\bibfnamefont {M.}~\bibnamefont
  {Veldhorst}}, \bibinfo {author} {\bibfnamefont {H.~G.~J.}\ \bibnamefont
  {Eenink}}, \bibinfo {author} {\bibfnamefont {C.~H.}\ \bibnamefont {Yang}},\
  and\ \bibinfo {author} {\bibfnamefont {A.~S.}\ \bibnamefont {Dzurak}},\
  }\bibfield  {title} {\bibinfo {title} {Silicon {CMOS} architecture for a
  spin-based quantum computer},\ }\href
  {https://doi.org/10.1038/s41467-017-01905-6} {\bibfield  {journal} {\bibinfo
  {journal} {Nature Communications}\ }\textbf {\bibinfo {volume} {8}},\
  \bibinfo {pages} {1766} (\bibinfo {year} {2017})}\BibitemShut {NoStop}%
\bibitem [{\citenamefont {Veldhorst}\ \emph {et~al.}(2015)\citenamefont
  {Veldhorst}, \citenamefont {Yang}, \citenamefont {Hwang}, \citenamefont
  {Huang}, \citenamefont {Dehollain}, \citenamefont {Muhonen}, \citenamefont
  {Simmons}, \citenamefont {Laucht}, \citenamefont {Hudson}, \citenamefont
  {Itoh}, \citenamefont {Morello},\ and\ \citenamefont
  {Dzurak}}]{VeldhorstNature15}%
  \BibitemOpen
  \bibfield  {author} {\bibinfo {author} {\bibfnamefont {M.}~\bibnamefont
  {Veldhorst}}, \bibinfo {author} {\bibfnamefont {C.~H.}\ \bibnamefont {Yang}},
  \bibinfo {author} {\bibfnamefont {J.~C.~C.}\ \bibnamefont {Hwang}}, \bibinfo
  {author} {\bibfnamefont {W.}~\bibnamefont {Huang}}, \bibinfo {author}
  {\bibfnamefont {J.~P.}\ \bibnamefont {Dehollain}}, \bibinfo {author}
  {\bibfnamefont {J.~T.}\ \bibnamefont {Muhonen}}, \bibinfo {author}
  {\bibfnamefont {S.}~\bibnamefont {Simmons}}, \bibinfo {author} {\bibfnamefont
  {A.}~\bibnamefont {Laucht}}, \bibinfo {author} {\bibfnamefont {F.~E.}\
  \bibnamefont {Hudson}}, \bibinfo {author} {\bibfnamefont {K.~M.}\
  \bibnamefont {Itoh}}, \bibinfo {author} {\bibfnamefont {A.}~\bibnamefont
  {Morello}},\ and\ \bibinfo {author} {\bibfnamefont {A.~S.}\ \bibnamefont
  {Dzurak}},\ }\bibfield  {title} {\bibinfo {title} {A two-qubit logic gate in
  silicon},\ }\href@noop {} {\bibfield  {journal} {\bibinfo  {journal}
  {Nature}\ }\textbf {\bibinfo {volume} {526}},\ \bibinfo {pages} {410}
  (\bibinfo {year} {2015})}\BibitemShut {NoStop}%
\bibitem [{\citenamefont {Hendrickx}\ \emph {et~al.}(2020)\citenamefont
  {Hendrickx}, \citenamefont {Franke}, \citenamefont {Sammak}, \citenamefont
  {Scappucci},\ and\ \citenamefont {Veldhorst}}]{Hendrickx2020}%
  \BibitemOpen
  \bibfield  {author} {\bibinfo {author} {\bibfnamefont {N.~W.}\ \bibnamefont
  {Hendrickx}}, \bibinfo {author} {\bibfnamefont {D.~P.}\ \bibnamefont
  {Franke}}, \bibinfo {author} {\bibfnamefont {A.}~\bibnamefont {Sammak}},
  \bibinfo {author} {\bibfnamefont {G.}~\bibnamefont {Scappucci}},\ and\
  \bibinfo {author} {\bibfnamefont {M.}~\bibnamefont {Veldhorst}},\ }\bibfield
  {title} {\bibinfo {title} {Fast two-qubit logic with holes in germanium},\
  }\href {https://doi.org/10.1038/s41586-019-1919-3} {\bibfield  {journal}
  {\bibinfo  {journal} {Nature}\ }\textbf {\bibinfo {volume} {577}},\ \bibinfo
  {pages} {487} (\bibinfo {year} {2020})}\BibitemShut {NoStop}%
\bibitem [{\citenamefont {Watson}\ \emph {et~al.}(2018)\citenamefont {Watson},
  \citenamefont {Philips}, \citenamefont {Kawakami}, \citenamefont {Ward},
  \citenamefont {Scarlino}, \citenamefont {Veldhorst}, \citenamefont {Savage},
  \citenamefont {Lagally}, \citenamefont {Friesen}, \citenamefont
  {Coppersmith}, \citenamefont {Eriksson},\ and\ \citenamefont
  {Vandersypen}}]{Watson2018}%
  \BibitemOpen
  \bibfield  {author} {\bibinfo {author} {\bibfnamefont {T.~F.}\ \bibnamefont
  {Watson}}, \bibinfo {author} {\bibfnamefont {S.~G.~J.}\ \bibnamefont
  {Philips}}, \bibinfo {author} {\bibfnamefont {E.}~\bibnamefont {Kawakami}},
  \bibinfo {author} {\bibfnamefont {D.~R.}\ \bibnamefont {Ward}}, \bibinfo
  {author} {\bibfnamefont {P.}~\bibnamefont {Scarlino}}, \bibinfo {author}
  {\bibfnamefont {M.}~\bibnamefont {Veldhorst}}, \bibinfo {author}
  {\bibfnamefont {D.~E.}\ \bibnamefont {Savage}}, \bibinfo {author}
  {\bibfnamefont {M.~G.}\ \bibnamefont {Lagally}}, \bibinfo {author}
  {\bibfnamefont {M.}~\bibnamefont {Friesen}}, \bibinfo {author} {\bibfnamefont
  {S.~N.}\ \bibnamefont {Coppersmith}}, \bibinfo {author} {\bibfnamefont
  {M.~A.}\ \bibnamefont {Eriksson}},\ and\ \bibinfo {author} {\bibfnamefont
  {L.~M.~K.}\ \bibnamefont {Vandersypen}},\ }\bibfield  {title} {\bibinfo
  {title} {A programmable two-qubit quantum processor in silicon},\ }\href
  {https://doi.org/10.1038/nature25766} {\bibfield  {journal} {\bibinfo
  {journal} {Nature}\ }\textbf {\bibinfo {volume} {555}},\ \bibinfo {pages}
  {633} (\bibinfo {year} {2018})}\BibitemShut {NoStop}%
\bibitem [{\citenamefont {He}\ \emph {et~al.}(2019)\citenamefont {He},
  \citenamefont {Gorman}, \citenamefont {Keith}, \citenamefont {Kranz},
  \citenamefont {Keizer},\ and\ \citenamefont {Simmons}}]{He2019_SWAP}%
  \BibitemOpen
  \bibfield  {author} {\bibinfo {author} {\bibfnamefont {Y.}~\bibnamefont
  {He}}, \bibinfo {author} {\bibfnamefont {S.~K.}\ \bibnamefont {Gorman}},
  \bibinfo {author} {\bibfnamefont {D.}~\bibnamefont {Keith}}, \bibinfo
  {author} {\bibfnamefont {L.}~\bibnamefont {Kranz}}, \bibinfo {author}
  {\bibfnamefont {J.~G.}\ \bibnamefont {Keizer}},\ and\ \bibinfo {author}
  {\bibfnamefont {M.~Y.}\ \bibnamefont {Simmons}},\ }\bibfield  {title}
  {\bibinfo {title} {A two-qubit gate between phosphorus donor electrons in
  silicon},\ }\href {https://doi.org/10.1038/s41586-019-1381-2} {\bibfield
  {journal} {\bibinfo  {journal} {Nature}\ }\textbf {\bibinfo {volume} {571}},\
  \bibinfo {pages} {371} (\bibinfo {year} {2019})}\BibitemShut {NoStop}%
\bibitem [{\citenamefont {Vandersypen}\ \emph {et~al.}(2017)\citenamefont
  {Vandersypen}, \citenamefont {Bluhm}, \citenamefont {Clarke}, \citenamefont
  {Dzurak}, \citenamefont {Ishihara}, \citenamefont {Morello}, \citenamefont
  {Reilly}, \citenamefont {Schreiber},\ and\ \citenamefont
  {Veldhorst}}]{VandersypenNPJ17}%
  \BibitemOpen
  \bibfield  {author} {\bibinfo {author} {\bibfnamefont {L.~M.~K.}\
  \bibnamefont {Vandersypen}}, \bibinfo {author} {\bibfnamefont
  {H.}~\bibnamefont {Bluhm}}, \bibinfo {author} {\bibfnamefont {J.~S.}\
  \bibnamefont {Clarke}}, \bibinfo {author} {\bibfnamefont {A.~S.}\
  \bibnamefont {Dzurak}}, \bibinfo {author} {\bibfnamefont {R.}~\bibnamefont
  {Ishihara}}, \bibinfo {author} {\bibfnamefont {A.}~\bibnamefont {Morello}},
  \bibinfo {author} {\bibfnamefont {D.~J.}\ \bibnamefont {Reilly}}, \bibinfo
  {author} {\bibfnamefont {L.~R.}\ \bibnamefont {Schreiber}},\ and\ \bibinfo
  {author} {\bibfnamefont {M.}~\bibnamefont {Veldhorst}},\ }\bibfield  {title}
  {\bibinfo {title} {Interfacing spin qubits in quantum dots and donors -- hot,
  dense, and coherent},\ }\href@noop {} {\bibfield  {journal} {\bibinfo
  {journal} {npj Quantum Information}\ }\textbf {\bibinfo {volume} {3}},\
  \bibinfo {pages} {34} (\bibinfo {year} {2017})}\BibitemShut {NoStop}%
\bibitem [{\citenamefont {Majer}\ \emph {et~al.}(2007)\citenamefont {Majer},
  \citenamefont {Chow}, \citenamefont {Gambetta}, \citenamefont {Koch},
  \citenamefont {Johnson}, \citenamefont {Schreier}, \citenamefont {Frunzio},
  \citenamefont {Schuster}, \citenamefont {Houck}, \citenamefont {Wallraff},
  \citenamefont {Blais}, \citenamefont {Devoret}, \citenamefont {Girvin},\ and\
  \citenamefont {Schoelkopf}}]{SchoelkopfNature07}%
  \BibitemOpen
  \bibfield  {author} {\bibinfo {author} {\bibfnamefont {J.}~\bibnamefont
  {Majer}}, \bibinfo {author} {\bibfnamefont {J.~M.}\ \bibnamefont {Chow}},
  \bibinfo {author} {\bibfnamefont {J.~M.}\ \bibnamefont {Gambetta}}, \bibinfo
  {author} {\bibfnamefont {J.}~\bibnamefont {Koch}}, \bibinfo {author}
  {\bibfnamefont {B.~R.}\ \bibnamefont {Johnson}}, \bibinfo {author}
  {\bibfnamefont {J.~A.}\ \bibnamefont {Schreier}}, \bibinfo {author}
  {\bibfnamefont {L.}~\bibnamefont {Frunzio}}, \bibinfo {author} {\bibfnamefont
  {D.~I.}\ \bibnamefont {Schuster}}, \bibinfo {author} {\bibfnamefont {A.~A.}\
  \bibnamefont {Houck}}, \bibinfo {author} {\bibfnamefont {A.}~\bibnamefont
  {Wallraff}}, \bibinfo {author} {\bibfnamefont {A.}~\bibnamefont {Blais}},
  \bibinfo {author} {\bibfnamefont {M.~H.}\ \bibnamefont {Devoret}}, \bibinfo
  {author} {\bibfnamefont {S.~M.}\ \bibnamefont {Girvin}},\ and\ \bibinfo
  {author} {\bibfnamefont {R.~J.}\ \bibnamefont {Schoelkopf}},\ }\bibfield
  {title} {\bibinfo {title} {Coupling superconducting qubits via a cavity
  bus},\ }\href@noop {} {\bibfield  {journal} {\bibinfo  {journal} {Nature}\
  }\textbf {\bibinfo {volume} {449}},\ \bibinfo {pages} {443} (\bibinfo {year}
  {2007})}\BibitemShut {NoStop}%
\bibitem [{\citenamefont {Sillanp{\"a}{\"a}}\ \emph {et~al.}(2007)\citenamefont
  {Sillanp{\"a}{\"a}}, \citenamefont {Park},\ and\ \citenamefont
  {Simmonds}}]{Simmonds07}%
  \BibitemOpen
  \bibfield  {author} {\bibinfo {author} {\bibfnamefont {M.~A.}\ \bibnamefont
  {Sillanp{\"a}{\"a}}}, \bibinfo {author} {\bibfnamefont {J.~I.}\ \bibnamefont
  {Park}},\ and\ \bibinfo {author} {\bibfnamefont {R.~W.}\ \bibnamefont
  {Simmonds}},\ }\bibfield  {title} {\bibinfo {title} {Coherent quantum state
  storage and transfer between two phase qubits via a resonant cavity},\
  }\href@noop {} {\bibfield  {journal} {\bibinfo  {journal} {Nature}\ }\textbf
  {\bibinfo {volume} {449}},\ \bibinfo {pages} {438} (\bibinfo {year}
  {2007})}\BibitemShut {NoStop}%
\bibitem [{\citenamefont {Benito}\ \emph {et~al.}(2019)\citenamefont {Benito},
  \citenamefont {Petta},\ and\ \citenamefont {Burkard}}]{Benito2qubit}%
  \BibitemOpen
  \bibfield  {author} {\bibinfo {author} {\bibfnamefont {M.}~\bibnamefont
  {Benito}}, \bibinfo {author} {\bibfnamefont {J.~R.}\ \bibnamefont {Petta}},\
  and\ \bibinfo {author} {\bibfnamefont {G.}~\bibnamefont {Burkard}},\
  }\bibfield  {title} {\bibinfo {title} {Optimized cavity-mediated dispersive
  two-qubit gates between spin qubits},\ }\href
  {https://doi.org/10.1103/PhysRevB.100.081412} {\bibfield  {journal} {\bibinfo
   {journal} {Phys. Rev. B}\ }\textbf {\bibinfo {volume} {100}},\ \bibinfo
  {pages} {081412} (\bibinfo {year} {2019})}\BibitemShut {NoStop}%
\bibitem [{\citenamefont {Hanson}\ \emph {et~al.}(2007)\citenamefont {Hanson},
  \citenamefont {Kouwenhoven}, \citenamefont {Petta}, \citenamefont {Tarucha},\
  and\ \citenamefont {Vandersypen}}]{Hanson2007}%
  \BibitemOpen
  \bibfield  {author} {\bibinfo {author} {\bibfnamefont {R.}~\bibnamefont
  {Hanson}}, \bibinfo {author} {\bibfnamefont {L.~P.}\ \bibnamefont
  {Kouwenhoven}}, \bibinfo {author} {\bibfnamefont {J.~R.}\ \bibnamefont
  {Petta}}, \bibinfo {author} {\bibfnamefont {S.}~\bibnamefont {Tarucha}},\
  and\ \bibinfo {author} {\bibfnamefont {L.~M.~K.}\ \bibnamefont
  {Vandersypen}},\ }\bibfield  {title} {\bibinfo {title} {Spins in few-electron
  quantum dots},\ }\href {https://doi.org/10.1103/RevModPhys.79.1217}
  {\bibfield  {journal} {\bibinfo  {journal} {Rev. Mod. Phys.}\ }\textbf
  {\bibinfo {volume} {79}},\ \bibinfo {pages} {1217} (\bibinfo {year}
  {2007})}\BibitemShut {NoStop}%
\bibitem [{\citenamefont {Schoelkopf}\ and\ \citenamefont
  {Girvin}(2008)}]{Schoelkopf2008}%
  \BibitemOpen
  \bibfield  {author} {\bibinfo {author} {\bibfnamefont {R.~J.}\ \bibnamefont
  {Schoelkopf}}\ and\ \bibinfo {author} {\bibfnamefont {S.~M.}\ \bibnamefont
  {Girvin}},\ }\bibfield  {title} {\bibinfo {title} {Wiring up quantum
  systems},\ }\href {https://doi.org/10.1038/451664a} {\bibfield  {journal}
  {\bibinfo  {journal} {Nature}\ }\textbf {\bibinfo {volume} {451}},\ \bibinfo
  {pages} {664} (\bibinfo {year} {2008})}\BibitemShut {NoStop}%
\bibitem [{\citenamefont {Golovach}\ \emph {et~al.}(2006)\citenamefont
  {Golovach}, \citenamefont {Borhani},\ and\ \citenamefont
  {Loss}}]{Golovach2006}%
  \BibitemOpen
  \bibfield  {author} {\bibinfo {author} {\bibfnamefont {V.~N.}\ \bibnamefont
  {Golovach}}, \bibinfo {author} {\bibfnamefont {M.}~\bibnamefont {Borhani}},\
  and\ \bibinfo {author} {\bibfnamefont {D.}~\bibnamefont {Loss}},\ }\bibfield
  {title} {\bibinfo {title} {Electric-dipole-induced spin resonance in quantum
  dots},\ }\href {https://doi.org/10.1103/PhysRevB.74.165319} {\bibfield
  {journal} {\bibinfo  {journal} {Phys. Rev. B}\ }\textbf {\bibinfo {volume}
  {74}},\ \bibinfo {pages} {165319} (\bibinfo {year} {2006})}\BibitemShut
  {NoStop}%
\bibitem [{\citenamefont {Pioro-Ladri\`{e}re}\ \emph
  {et~al.}(2008)\citenamefont {Pioro-Ladri\`{e}re}, \citenamefont {Obata},
  \citenamefont {Tokura}, \citenamefont {Shin}, \citenamefont {Kubo},
  \citenamefont {Yoshida}, \citenamefont {Taniyama},\ and\ \citenamefont
  {Tarucha}}]{Pioro-Ladriere2008}%
  \BibitemOpen
  \bibfield  {author} {\bibinfo {author} {\bibfnamefont {M.}~\bibnamefont
  {Pioro-Ladri\`{e}re}}, \bibinfo {author} {\bibfnamefont {T.}~\bibnamefont
  {Obata}}, \bibinfo {author} {\bibfnamefont {Y.}~\bibnamefont {Tokura}},
  \bibinfo {author} {\bibfnamefont {Y.-S.}\ \bibnamefont {Shin}}, \bibinfo
  {author} {\bibfnamefont {T.}~\bibnamefont {Kubo}}, \bibinfo {author}
  {\bibfnamefont {K.}~\bibnamefont {Yoshida}}, \bibinfo {author} {\bibfnamefont
  {T.}~\bibnamefont {Taniyama}},\ and\ \bibinfo {author} {\bibfnamefont
  {S.}~\bibnamefont {Tarucha}},\ }\bibfield  {title} {\bibinfo {title}
  {Electrically driven single-electron spin resonance in a slanting {Z}eeman
  field},\ }\href {https://doi.org/10.1038/nphys1053} {\bibfield  {journal}
  {\bibinfo  {journal} {Nature Physics}\ }\textbf {\bibinfo {volume} {4}},\
  \bibinfo {pages} {776} (\bibinfo {year} {2008})}\BibitemShut {NoStop}%
\bibitem [{\citenamefont {Nadj-Perge}\ \emph {et~al.}(2010)\citenamefont
  {Nadj-Perge}, \citenamefont {Frolov}, \citenamefont {Bakkers},\ and\
  \citenamefont {Kouwenhoven}}]{Nadj-Perge2010}%
  \BibitemOpen
  \bibfield  {author} {\bibinfo {author} {\bibfnamefont {S.}~\bibnamefont
  {Nadj-Perge}}, \bibinfo {author} {\bibfnamefont {S.~M.}\ \bibnamefont
  {Frolov}}, \bibinfo {author} {\bibfnamefont {E.~P. A.~M.}\ \bibnamefont
  {Bakkers}},\ and\ \bibinfo {author} {\bibfnamefont {L.~P.}\ \bibnamefont
  {Kouwenhoven}},\ }\bibfield  {title} {\bibinfo {title} {Spin-orbit qubit in a
  semiconductor nanowire},\ }\href {https://doi.org/10.1038/nature09682}
  {\bibfield  {journal} {\bibinfo  {journal} {Nature}\ }\textbf {\bibinfo
  {volume} {468}},\ \bibinfo {pages} {1084} (\bibinfo {year}
  {2010})}\BibitemShut {NoStop}%
\bibitem [{\citenamefont {Petersson}\ \emph {et~al.}(2012)\citenamefont
  {Petersson}, \citenamefont {McFaul}, \citenamefont {Schroer}, \citenamefont
  {Jung}, \citenamefont {Taylor}, \citenamefont {Houck},\ and\ \citenamefont
  {Petta}}]{Petersson2012}%
  \BibitemOpen
  \bibfield  {author} {\bibinfo {author} {\bibfnamefont {K.~D.}\ \bibnamefont
  {Petersson}}, \bibinfo {author} {\bibfnamefont {L.~W.}\ \bibnamefont
  {McFaul}}, \bibinfo {author} {\bibfnamefont {M.~D.}\ \bibnamefont {Schroer}},
  \bibinfo {author} {\bibfnamefont {M.}~\bibnamefont {Jung}}, \bibinfo {author}
  {\bibfnamefont {J.~M.}\ \bibnamefont {Taylor}}, \bibinfo {author}
  {\bibfnamefont {A.~A.}\ \bibnamefont {Houck}},\ and\ \bibinfo {author}
  {\bibfnamefont {J.~R.}\ \bibnamefont {Petta}},\ }\bibfield  {title} {\bibinfo
  {title} {Circuit quantum electrodynamics with a spin qubit},\ }\href
  {https://doi.org/10.1038/nature11559} {\bibfield  {journal} {\bibinfo
  {journal} {Nature}\ }\textbf {\bibinfo {volume} {490}},\ \bibinfo {pages}
  {380} (\bibinfo {year} {2012})}\BibitemShut {NoStop}%
\bibitem [{\citenamefont {Salfi}\ \emph {et~al.}(2016)\citenamefont {Salfi},
  \citenamefont {Mol}, \citenamefont {Culcer},\ and\ \citenamefont
  {Rogge}}]{Salfi2016}%
  \BibitemOpen
  \bibfield  {author} {\bibinfo {author} {\bibfnamefont {J.}~\bibnamefont
  {Salfi}}, \bibinfo {author} {\bibfnamefont {J.~A.}\ \bibnamefont {Mol}},
  \bibinfo {author} {\bibfnamefont {D.}~\bibnamefont {Culcer}},\ and\ \bibinfo
  {author} {\bibfnamefont {S.}~\bibnamefont {Rogge}},\ }\bibfield  {title}
  {\bibinfo {title} {Charge-{I}nsensitive {S}ingle-{A}tom {S}pin-{O}rbit
  {Q}ubit in {S}ilicon},\ }\href
  {https://doi.org/10.1103/PhysRevLett.116.246801} {\bibfield  {journal}
  {\bibinfo  {journal} {Phys. Rev. Lett.}\ }\textbf {\bibinfo {volume} {116}},\
  \bibinfo {pages} {246801} (\bibinfo {year} {2016})}\BibitemShut {NoStop}%
\bibitem [{\citenamefont {Corna}\ \emph {et~al.}(2018)\citenamefont {Corna},
  \citenamefont {Bourdet}, \citenamefont {Maurand}, \citenamefont {Crippa},
  \citenamefont {Kotekar-Patil}, \citenamefont {Bohuslavskyi}, \citenamefont
  {Lavi{\'e}ville}, \citenamefont {Hutin}, \citenamefont {Barraud},
  \citenamefont {Jehl}, \citenamefont {Vinet}, \citenamefont {De~Franceschi},
  \citenamefont {Niquet},\ and\ \citenamefont {Sanquer}}]{Corna2018}%
  \BibitemOpen
  \bibfield  {author} {\bibinfo {author} {\bibfnamefont {A.}~\bibnamefont
  {Corna}}, \bibinfo {author} {\bibfnamefont {L.}~\bibnamefont {Bourdet}},
  \bibinfo {author} {\bibfnamefont {R.}~\bibnamefont {Maurand}}, \bibinfo
  {author} {\bibfnamefont {A.}~\bibnamefont {Crippa}}, \bibinfo {author}
  {\bibfnamefont {D.}~\bibnamefont {Kotekar-Patil}}, \bibinfo {author}
  {\bibfnamefont {H.}~\bibnamefont {Bohuslavskyi}}, \bibinfo {author}
  {\bibfnamefont {R.}~\bibnamefont {Lavi{\'e}ville}}, \bibinfo {author}
  {\bibfnamefont {L.}~\bibnamefont {Hutin}}, \bibinfo {author} {\bibfnamefont
  {S.}~\bibnamefont {Barraud}}, \bibinfo {author} {\bibfnamefont
  {X.}~\bibnamefont {Jehl}}, \bibinfo {author} {\bibfnamefont {M.}~\bibnamefont
  {Vinet}}, \bibinfo {author} {\bibfnamefont {S.}~\bibnamefont
  {De~Franceschi}}, \bibinfo {author} {\bibfnamefont {Y.-M.}\ \bibnamefont
  {Niquet}},\ and\ \bibinfo {author} {\bibfnamefont {M.}~\bibnamefont
  {Sanquer}},\ }\bibfield  {title} {\bibinfo {title} {Electrically driven
  electron spin resonance mediated by spin-valley-orbit coupling in a silicon
  quantum dot},\ }\href {https://doi.org/10.1038/s41534-018-0059-1} {\bibfield
  {journal} {\bibinfo  {journal} {npj Quantum Information}\ }\textbf {\bibinfo
  {volume} {4}},\ \bibinfo {pages} {6} (\bibinfo {year} {2018})}\BibitemShut
  {NoStop}%
\bibitem [{\citenamefont {Watzinger}\ \emph {et~al.}(2018)\citenamefont
  {Watzinger}, \citenamefont {Kuku\v{c}ka}, \citenamefont {Vuku\v{s}i\'{c}},
  \citenamefont {Gao}, \citenamefont {Wang}, \citenamefont {Sch\"{a}ffler},
  \citenamefont {Zhang},\ and\ \citenamefont {Katsaros}}]{Watzinger2018}%
  \BibitemOpen
  \bibfield  {author} {\bibinfo {author} {\bibfnamefont {H.}~\bibnamefont
  {Watzinger}}, \bibinfo {author} {\bibfnamefont {J.}~\bibnamefont
  {Kuku\v{c}ka}}, \bibinfo {author} {\bibfnamefont {L.}~\bibnamefont
  {Vuku\v{s}i\'{c}}}, \bibinfo {author} {\bibfnamefont {F.}~\bibnamefont
  {Gao}}, \bibinfo {author} {\bibfnamefont {T.}~\bibnamefont {Wang}}, \bibinfo
  {author} {\bibfnamefont {F.}~\bibnamefont {Sch\"{a}ffler}}, \bibinfo {author}
  {\bibfnamefont {J.-J.}\ \bibnamefont {Zhang}},\ and\ \bibinfo {author}
  {\bibfnamefont {G.}~\bibnamefont {Katsaros}},\ }\bibfield  {title} {\bibinfo
  {title} {A germanium hole spin qubit},\ }\href
  {https://doi.org/10.1038/s41467-018-06418-4} {\bibfield  {journal} {\bibinfo
  {journal} {Nature Communications}\ }\textbf {\bibinfo {volume} {9}},\
  \bibinfo {pages} {3902} (\bibinfo {year} {2018})}\BibitemShut {NoStop}%
\bibitem [{\citenamefont {Wang}\ \emph {et~al.}()\citenamefont {Wang},
  \citenamefont {Chen}, \citenamefont {Klimeck}, \citenamefont {Simmons},\ and\
  \citenamefont {Rahman}}]{arxiv1P2P}%
  \BibitemOpen
  \bibfield  {author} {\bibinfo {author} {\bibfnamefont {Y.}~\bibnamefont
  {Wang}}, \bibinfo {author} {\bibfnamefont {C.-Y.}\ \bibnamefont {Chen}},
  \bibinfo {author} {\bibfnamefont {G.}~\bibnamefont {Klimeck}}, \bibinfo
  {author} {\bibfnamefont {M.~Y.}\ \bibnamefont {Simmons}},\ and\ \bibinfo
  {author} {\bibfnamefont {R.}~\bibnamefont {Rahman}},\ }\bibfield  {title}
  {\bibinfo {title} {All-electrical control of donor-bound electron spin qubits
  in silicon},\ }\href@noop {} {\bibinfo  {journal} {arXiv:1703.05370 (2017)}\
  }\BibitemShut {NoStop}%
\bibitem [{\citenamefont {Krauth}\ \emph {et~al.}()\citenamefont {Krauth},
  \citenamefont {Gorman}, \citenamefont {He}, \citenamefont {Jones},
  \citenamefont {Macha}, \citenamefont {Kocsis}, \citenamefont {Chua},
  \citenamefont {Voisin}, \citenamefont {Rogge}, \citenamefont {Rahman},
  \citenamefont {Chung},\ and\ \citenamefont {Simmons}}]{Felix2021}%
  \BibitemOpen
\bibfield  {journal} {  }\bibfield  {author} {\bibinfo {author} {\bibfnamefont
  {F.~N.}\ \bibnamefont {Krauth}}, \bibinfo {author} {\bibfnamefont {S.~K.}\
  \bibnamefont {Gorman}}, \bibinfo {author} {\bibfnamefont {Y.}~\bibnamefont
  {He}}, \bibinfo {author} {\bibfnamefont {M.~T.}\ \bibnamefont {Jones}},
  \bibinfo {author} {\bibfnamefont {P.}~\bibnamefont {Macha}}, \bibinfo
  {author} {\bibfnamefont {S.}~\bibnamefont {Kocsis}}, \bibinfo {author}
  {\bibfnamefont {C.}~\bibnamefont {Chua}}, \bibinfo {author} {\bibfnamefont
  {B.}~\bibnamefont {Voisin}}, \bibinfo {author} {\bibfnamefont
  {S.}~\bibnamefont {Rogge}}, \bibinfo {author} {\bibfnamefont
  {R.}~\bibnamefont {Rahman}}, \bibinfo {author} {\bibfnamefont
  {Y.}~\bibnamefont {Chung}},\ and\ \bibinfo {author} {\bibfnamefont {M.~Y.}\
  \bibnamefont {Simmons}},\ }\href@noop {} {\bibinfo  {journal} {Unpublished}\
  }\BibitemShut {NoStop}%
\bibitem [{\citenamefont {Burkard}\ \emph {et~al.}(2020)\citenamefont
  {Burkard}, \citenamefont {Gullans}, \citenamefont {Mi},\ and\ \citenamefont
  {Petta}}]{Burkard2020}%
  \BibitemOpen
\bibfield  {journal} {  }\bibfield  {author} {\bibinfo {author} {\bibfnamefont
  {G.}~\bibnamefont {Burkard}}, \bibinfo {author} {\bibfnamefont {M.~J.}\
  \bibnamefont {Gullans}}, \bibinfo {author} {\bibfnamefont {X.}~\bibnamefont
  {Mi}},\ and\ \bibinfo {author} {\bibfnamefont {J.~R.}\ \bibnamefont
  {Petta}},\ }\bibfield  {title} {\bibinfo {title}
  {Superconductor--semiconductor hybrid-circuit quantum electrodynamics},\
  }\href {https://doi.org/10.1038/s42254-019-0135-2} {\bibfield  {journal}
  {\bibinfo  {journal} {Nature Reviews Physics}\ }\textbf {\bibinfo {volume}
  {2}},\ \bibinfo {pages} {129} (\bibinfo {year} {2020})}\BibitemShut {NoStop}%
\bibitem [{\citenamefont {Burkard}\ and\ \citenamefont
  {Imamoglu}(2006)}]{Burkard2006}%
  \BibitemOpen
  \bibfield  {author} {\bibinfo {author} {\bibfnamefont {G.}~\bibnamefont
  {Burkard}}\ and\ \bibinfo {author} {\bibfnamefont {A.}~\bibnamefont
  {Imamoglu}},\ }\bibfield  {title} {\bibinfo {title} {Ultra-long-distance
  interaction between spin qubits},\ }\href
  {https://doi.org/10.1103/PhysRevB.74.041307} {\bibfield  {journal} {\bibinfo
  {journal} {Phys. Rev. B}\ }\textbf {\bibinfo {volume} {74}},\ \bibinfo
  {pages} {041307} (\bibinfo {year} {2006})}\BibitemShut {NoStop}%
\bibitem [{\citenamefont {Beaudoin}\ \emph {et~al.}(2017)\citenamefont
  {Beaudoin}, \citenamefont {Lachance-Quirion}, \citenamefont {Coish},\ and\
  \citenamefont {Pioro-Ladri\`{e}re}}]{PioroNano17}%
  \BibitemOpen
  \bibfield  {author} {\bibinfo {author} {\bibfnamefont {F.}~\bibnamefont
  {Beaudoin}}, \bibinfo {author} {\bibfnamefont {D.}~\bibnamefont
  {Lachance-Quirion}}, \bibinfo {author} {\bibfnamefont {W.~A.}\ \bibnamefont
  {Coish}},\ and\ \bibinfo {author} {\bibfnamefont {M.}~\bibnamefont
  {Pioro-Ladri\`{e}re}},\ }\bibfield  {title} {\bibinfo {title} {Coupling a
  single electron spin to a microwave resonator: controlling transverse and
  longitudinal couplings},\ }\href@noop {} {\bibfield  {journal} {\bibinfo
  {journal} {Nanotechnology}\ }\textbf {\bibinfo {volume} {27}},\ \bibinfo
  {pages} {464003} (\bibinfo {year} {2017})}\BibitemShut {NoStop}%
\bibitem [{\citenamefont {Hu}\ \emph {et~al.}(2012)\citenamefont {Hu},
  \citenamefont {Liu},\ and\ \citenamefont {Nori}}]{Hu2012}%
  \BibitemOpen
  \bibfield  {author} {\bibinfo {author} {\bibfnamefont {X.}~\bibnamefont
  {Hu}}, \bibinfo {author} {\bibfnamefont {Y.-x.}\ \bibnamefont {Liu}},\ and\
  \bibinfo {author} {\bibfnamefont {F.}~\bibnamefont {Nori}},\ }\bibfield
  {title} {\bibinfo {title} {Strong coupling of a spin qubit to a
  superconducting stripline cavity},\ }\href
  {https://doi.org/10.1103/PhysRevB.86.035314} {\bibfield  {journal} {\bibinfo
  {journal} {Phys. Rev. B}\ }\textbf {\bibinfo {volume} {86}},\ \bibinfo
  {pages} {035314} (\bibinfo {year} {2012})}\BibitemShut {NoStop}%
\bibitem [{\citenamefont {Mi}\ \emph {et~al.}(2018)\citenamefont {Mi},
  \citenamefont {Benito}, \citenamefont {Putz}, \citenamefont {Zajac},
  \citenamefont {Taylor}, \citenamefont {Burkard},\ and\ \citenamefont
  {Petta}}]{Mi2018}%
  \BibitemOpen
  \bibfield  {author} {\bibinfo {author} {\bibfnamefont {X.}~\bibnamefont
  {Mi}}, \bibinfo {author} {\bibfnamefont {M.}~\bibnamefont {Benito}}, \bibinfo
  {author} {\bibfnamefont {S.}~\bibnamefont {Putz}}, \bibinfo {author}
  {\bibfnamefont {D.~M.}\ \bibnamefont {Zajac}}, \bibinfo {author}
  {\bibfnamefont {J.~M.}\ \bibnamefont {Taylor}}, \bibinfo {author}
  {\bibfnamefont {G.}~\bibnamefont {Burkard}},\ and\ \bibinfo {author}
  {\bibfnamefont {J.~R.}\ \bibnamefont {Petta}},\ }\bibfield  {title} {\bibinfo
  {title} {A coherent spin--photon interface in silicon},\ }\href
  {https://doi.org/10.1038/nature25769} {\bibfield  {journal} {\bibinfo
  {journal} {Nature}\ }\textbf {\bibinfo {volume} {555}},\ \bibinfo {pages}
  {599} (\bibinfo {year} {2018})}\BibitemShut {NoStop}%
\bibitem [{\citenamefont {Samkharadze}\ \emph {et~al.}(2018)\citenamefont
  {Samkharadze}, \citenamefont {Zheng}, \citenamefont {Kalhor}, \citenamefont
  {Brousse}, \citenamefont {Sammak}, \citenamefont {Mendes}, \citenamefont
  {Blais}, \citenamefont {Scappucci},\ and\ \citenamefont
  {Vandersypen}}]{Samkharadze2018}%
  \BibitemOpen
  \bibfield  {author} {\bibinfo {author} {\bibfnamefont {N.}~\bibnamefont
  {Samkharadze}}, \bibinfo {author} {\bibfnamefont {G.}~\bibnamefont {Zheng}},
  \bibinfo {author} {\bibfnamefont {N.}~\bibnamefont {Kalhor}}, \bibinfo
  {author} {\bibfnamefont {D.}~\bibnamefont {Brousse}}, \bibinfo {author}
  {\bibfnamefont {A.}~\bibnamefont {Sammak}}, \bibinfo {author} {\bibfnamefont
  {U.~C.}\ \bibnamefont {Mendes}}, \bibinfo {author} {\bibfnamefont
  {A.}~\bibnamefont {Blais}}, \bibinfo {author} {\bibfnamefont
  {G.}~\bibnamefont {Scappucci}},\ and\ \bibinfo {author} {\bibfnamefont
  {L.~M.~K.}\ \bibnamefont {Vandersypen}},\ }\bibfield  {title} {\bibinfo
  {title} {Strong spin-photon coupling in silicon},\ }\href
  {https://doi.org/10.1126/science.aar4054} {\bibfield  {journal} {\bibinfo
  {journal} {Science}\ }\textbf {\bibinfo {volume} {359}},\ \bibinfo {pages}
  {1123} (\bibinfo {year} {2018})}\BibitemShut {NoStop}%
\bibitem [{\citenamefont {Borjans}\ \emph {et~al.}(2019)\citenamefont
  {Borjans}, \citenamefont {Croot}, \citenamefont {Mi}, \citenamefont
  {Gullans},\ and\ \citenamefont {Petta}}]{PettaNature19}%
  \BibitemOpen
  \bibfield  {author} {\bibinfo {author} {\bibfnamefont {F.}~\bibnamefont
  {Borjans}}, \bibinfo {author} {\bibfnamefont {X.~G.}\ \bibnamefont {Croot}},
  \bibinfo {author} {\bibfnamefont {X.}~\bibnamefont {Mi}}, \bibinfo {author}
  {\bibfnamefont {M.~J.}\ \bibnamefont {Gullans}},\ and\ \bibinfo {author}
  {\bibfnamefont {J.~R.}\ \bibnamefont {Petta}},\ }\bibfield  {title} {\bibinfo
  {title} {Resonant microwave-mediated interactions between distant electron
  spins},\ }\href@noop {} {\bibfield  {journal} {\bibinfo  {journal} {Nature}\
  }\textbf {\bibinfo {volume} {577}},\ \bibinfo {pages} {195} (\bibinfo {year}
  {2019})}\BibitemShut {NoStop}%
\bibitem [{\citenamefont {Landig}\ \emph {et~al.}(2018)\citenamefont {Landig},
  \citenamefont {Koski}, \citenamefont {Scarlino}, \citenamefont {Mendes},
  \citenamefont {Blais}, \citenamefont {Reichl}, \citenamefont {Wegscheider},
  \citenamefont {Wallraff}, \citenamefont {Ensslin},\ and\ \citenamefont
  {Ihn}}]{Landig2018}%
  \BibitemOpen
  \bibfield  {author} {\bibinfo {author} {\bibfnamefont {A.~J.}\ \bibnamefont
  {Landig}}, \bibinfo {author} {\bibfnamefont {J.~V.}\ \bibnamefont {Koski}},
  \bibinfo {author} {\bibfnamefont {P.}~\bibnamefont {Scarlino}}, \bibinfo
  {author} {\bibfnamefont {U.~C.}\ \bibnamefont {Mendes}}, \bibinfo {author}
  {\bibfnamefont {A.}~\bibnamefont {Blais}}, \bibinfo {author} {\bibfnamefont
  {C.}~\bibnamefont {Reichl}}, \bibinfo {author} {\bibfnamefont
  {W.}~\bibnamefont {Wegscheider}}, \bibinfo {author} {\bibfnamefont
  {A.}~\bibnamefont {Wallraff}}, \bibinfo {author} {\bibfnamefont
  {K.}~\bibnamefont {Ensslin}},\ and\ \bibinfo {author} {\bibfnamefont
  {T.}~\bibnamefont {Ihn}},\ }\bibfield  {title} {\bibinfo {title} {Coherent
  spin-photon coupling using a resonant exchange qubit},\ }\href
  {https://doi.org/10.1038/s41586-018-0365-y} {\bibfield  {journal} {\bibinfo
  {journal} {Nature}\ }\textbf {\bibinfo {volume} {560}},\ \bibinfo {pages}
  {179} (\bibinfo {year} {2018})}\BibitemShut {NoStop}%
\bibitem [{\citenamefont {Landig}\ \emph {et~al.}(2019)\citenamefont {Landig},
  \citenamefont {Koski}, \citenamefont {Scarlino}, \citenamefont {M\"{u}ller},
  \citenamefont {Abadillo-Uriel}, \citenamefont {Kratochwil}, \citenamefont
  {Reichl}, \citenamefont {Wegscheider}, \citenamefont {Coppersmith},
  \citenamefont {Friesen}, \citenamefont {Wallraff}, \citenamefont {Ihn},\ and\
  \citenamefont {Ensslin}}]{WallraffNatComms19}%
  \BibitemOpen
  \bibfield  {author} {\bibinfo {author} {\bibfnamefont {A.~J.}\ \bibnamefont
  {Landig}}, \bibinfo {author} {\bibfnamefont {J.~V.}\ \bibnamefont {Koski}},
  \bibinfo {author} {\bibfnamefont {P.}~\bibnamefont {Scarlino}}, \bibinfo
  {author} {\bibfnamefont {C.}~\bibnamefont {M\"{u}ller}}, \bibinfo {author}
  {\bibfnamefont {J.~C.}\ \bibnamefont {Abadillo-Uriel}}, \bibinfo {author}
  {\bibfnamefont {B.}~\bibnamefont {Kratochwil}}, \bibinfo {author}
  {\bibfnamefont {C.}~\bibnamefont {Reichl}}, \bibinfo {author} {\bibfnamefont
  {W.}~\bibnamefont {Wegscheider}}, \bibinfo {author} {\bibfnamefont {S.~N.}\
  \bibnamefont {Coppersmith}}, \bibinfo {author} {\bibfnamefont
  {M.}~\bibnamefont {Friesen}}, \bibinfo {author} {\bibfnamefont
  {A.}~\bibnamefont {Wallraff}}, \bibinfo {author} {\bibfnamefont
  {T.}~\bibnamefont {Ihn}},\ and\ \bibinfo {author} {\bibfnamefont
  {K.}~\bibnamefont {Ensslin}},\ }\bibfield  {title} {\bibinfo {title}
  {Virtual-photon-mediated spin-qubit-transmon coupling},\ }\href@noop {}
  {\bibfield  {journal} {\bibinfo  {journal} {Nature Communications}\ }\textbf
  {\bibinfo {volume} {10}},\ \bibinfo {pages} {5037} (\bibinfo {year}
  {2019})}\BibitemShut {NoStop}%
\bibitem [{\citenamefont {Koski}\ \emph {et~al.}(2020)\citenamefont {Koski},
  \citenamefont {Landig}, \citenamefont {Russ}, \citenamefont {Abadillo-Uriel},
  \citenamefont {Scarlino}, \citenamefont {Kratochwil}, \citenamefont {Reichl},
  \citenamefont {Wegscheider}, \citenamefont {Burkard}, \citenamefont
  {Friesen}, \citenamefont {Coppersmith}, \citenamefont {Wallraff},
  \citenamefont {Ensslin},\ and\ \citenamefont {Ihn}}]{Koski2020}%
  \BibitemOpen
  \bibfield  {author} {\bibinfo {author} {\bibfnamefont {J.~V.}\ \bibnamefont
  {Koski}}, \bibinfo {author} {\bibfnamefont {A.~J.}\ \bibnamefont {Landig}},
  \bibinfo {author} {\bibfnamefont {M.}~\bibnamefont {Russ}}, \bibinfo {author}
  {\bibfnamefont {J.~C.}\ \bibnamefont {Abadillo-Uriel}}, \bibinfo {author}
  {\bibfnamefont {P.}~\bibnamefont {Scarlino}}, \bibinfo {author}
  {\bibfnamefont {B.}~\bibnamefont {Kratochwil}}, \bibinfo {author}
  {\bibfnamefont {C.}~\bibnamefont {Reichl}}, \bibinfo {author} {\bibfnamefont
  {W.}~\bibnamefont {Wegscheider}}, \bibinfo {author} {\bibfnamefont
  {G.}~\bibnamefont {Burkard}}, \bibinfo {author} {\bibfnamefont
  {M.}~\bibnamefont {Friesen}}, \bibinfo {author} {\bibfnamefont {S.~N.}\
  \bibnamefont {Coppersmith}}, \bibinfo {author} {\bibfnamefont
  {A.}~\bibnamefont {Wallraff}}, \bibinfo {author} {\bibfnamefont
  {K.}~\bibnamefont {Ensslin}},\ and\ \bibinfo {author} {\bibfnamefont
  {T.}~\bibnamefont {Ihn}},\ }\bibfield  {title} {\bibinfo {title} {Strong
  photon coupling to the quadrupole moment of an electron in a solid-state
  qubit},\ }\href {https://doi.org/10.1038/s41567-020-0862-4} {\bibfield
  {journal} {\bibinfo  {journal} {Nature Physics}\ }\textbf {\bibinfo {volume}
  {16}},\ \bibinfo {pages} {642} (\bibinfo {year} {2020})}\BibitemShut
  {NoStop}%
\bibitem [{\citenamefont {Muhonen}\ \emph {et~al.}(2014)\citenamefont
  {Muhonen}, \citenamefont {Dehollain}, \citenamefont {Laucht}, \citenamefont
  {Hudson}, \citenamefont {Kalra}, \citenamefont {Sekiguchi}, \citenamefont
  {Itoh}, \citenamefont {Jamieson}, \citenamefont {McCallum}, \citenamefont
  {Dzurak},\ and\ \citenamefont {Morello}}]{Muhonen2014}%
  \BibitemOpen
  \bibfield  {author} {\bibinfo {author} {\bibfnamefont {J.~T.}\ \bibnamefont
  {Muhonen}}, \bibinfo {author} {\bibfnamefont {J.~P.}\ \bibnamefont
  {Dehollain}}, \bibinfo {author} {\bibfnamefont {A.}~\bibnamefont {Laucht}},
  \bibinfo {author} {\bibfnamefont {F.~E.}\ \bibnamefont {Hudson}}, \bibinfo
  {author} {\bibfnamefont {R.}~\bibnamefont {Kalra}}, \bibinfo {author}
  {\bibfnamefont {T.}~\bibnamefont {Sekiguchi}}, \bibinfo {author}
  {\bibfnamefont {K.~M.}\ \bibnamefont {Itoh}}, \bibinfo {author}
  {\bibfnamefont {D.~N.}\ \bibnamefont {Jamieson}}, \bibinfo {author}
  {\bibfnamefont {J.~C.}\ \bibnamefont {McCallum}}, \bibinfo {author}
  {\bibfnamefont {A.~S.}\ \bibnamefont {Dzurak}},\ and\ \bibinfo {author}
  {\bibfnamefont {A.}~\bibnamefont {Morello}},\ }\bibfield  {title} {\bibinfo
  {title} {Storing quantum information for 30 seconds in a nanoelectronic
  device},\ }\href {https://doi.org/10.1038/nnano.2014.211} {\bibfield
  {journal} {\bibinfo  {journal} {Nature Nanotechnology}\ }\textbf {\bibinfo
  {volume} {9}},\ \bibinfo {pages} {986} (\bibinfo {year} {2014})}\BibitemShut
  {NoStop}%
\bibitem [{\citenamefont {Watson}\ \emph {et~al.}(2017)\citenamefont {Watson},
  \citenamefont {Weber}, \citenamefont {Hsueh}, \citenamefont {Hollenberg},
  \citenamefont {Rahman},\ and\ \citenamefont {Simmons}}]{Watson2017}%
  \BibitemOpen
  \bibfield  {author} {\bibinfo {author} {\bibfnamefont {T.~F.}\ \bibnamefont
  {Watson}}, \bibinfo {author} {\bibfnamefont {B.}~\bibnamefont {Weber}},
  \bibinfo {author} {\bibfnamefont {Y.-L.}\ \bibnamefont {Hsueh}}, \bibinfo
  {author} {\bibfnamefont {L.~C.~L.}\ \bibnamefont {Hollenberg}}, \bibinfo
  {author} {\bibfnamefont {R.}~\bibnamefont {Rahman}},\ and\ \bibinfo {author}
  {\bibfnamefont {M.~Y.}\ \bibnamefont {Simmons}},\ }\bibfield  {title}
  {\bibinfo {title} {Atomically engineered electron spin lifetimes of 30 s in
  silicon},\ }\href {https://doi.org/10.1126/sciadv.1602811} {\bibfield
  {journal} {\bibinfo  {journal} {Science Advances}\ }\textbf {\bibinfo
  {volume} {3}},\ \bibinfo {pages} {e1602811} (\bibinfo {year}
  {2017})}\BibitemShut {NoStop}%
\bibitem [{\citenamefont {Fuechsle}\ \emph {et~al.}(2012)\citenamefont
  {Fuechsle}, \citenamefont {Miwa}, \citenamefont {Mahapatra}, \citenamefont
  {Ryu}, \citenamefont {Lee}, \citenamefont {Warschkow}, \citenamefont
  {Hollenberg}, \citenamefont {Klimeck},\ and\ \citenamefont
  {Simmons}}]{Fuechsle2012}%
  \BibitemOpen
  \bibfield  {author} {\bibinfo {author} {\bibfnamefont {M.}~\bibnamefont
  {Fuechsle}}, \bibinfo {author} {\bibfnamefont {J.~A.}\ \bibnamefont {Miwa}},
  \bibinfo {author} {\bibfnamefont {S.}~\bibnamefont {Mahapatra}}, \bibinfo
  {author} {\bibfnamefont {H.}~\bibnamefont {Ryu}}, \bibinfo {author}
  {\bibfnamefont {S.}~\bibnamefont {Lee}}, \bibinfo {author} {\bibfnamefont
  {O.}~\bibnamefont {Warschkow}}, \bibinfo {author} {\bibfnamefont {L.~C.~L.}\
  \bibnamefont {Hollenberg}}, \bibinfo {author} {\bibfnamefont
  {G.}~\bibnamefont {Klimeck}},\ and\ \bibinfo {author} {\bibfnamefont {M.~Y.}\
  \bibnamefont {Simmons}},\ }\bibfield  {title} {\bibinfo {title} {A
  single-atom transistor},\ }\href {https://doi.org/10.1038/nnano.2012.21}
  {\bibfield  {journal} {\bibinfo  {journal} {Nature Nanotechnology}\ }\textbf
  {\bibinfo {volume} {7}},\ \bibinfo {pages} {242} (\bibinfo {year}
  {2012})}\BibitemShut {NoStop}%
\bibitem [{\citenamefont {Koch}\ \emph {et~al.}(2019)\citenamefont {Koch},
  \citenamefont {Keizer}, \citenamefont {Pakkiam}, \citenamefont {Keith},
  \citenamefont {House}, \citenamefont {Peretz},\ and\ \citenamefont
  {Simmons}}]{KochNatureNano19}%
  \BibitemOpen
  \bibfield  {author} {\bibinfo {author} {\bibfnamefont {M.}~\bibnamefont
  {Koch}}, \bibinfo {author} {\bibfnamefont {J.~G.}\ \bibnamefont {Keizer}},
  \bibinfo {author} {\bibfnamefont {P.}~\bibnamefont {Pakkiam}}, \bibinfo
  {author} {\bibfnamefont {D.}~\bibnamefont {Keith}}, \bibinfo {author}
  {\bibfnamefont {M.~G.}\ \bibnamefont {House}}, \bibinfo {author}
  {\bibfnamefont {E.}~\bibnamefont {Peretz}},\ and\ \bibinfo {author}
  {\bibfnamefont {M.~Y.}\ \bibnamefont {Simmons}},\ }\bibfield  {title}
  {\bibinfo {title} {Spin read-out in atomic qubits in an all-epitaxial
  three-dimensional transistor},\ }\href@noop {} {\bibfield  {journal}
  {\bibinfo  {journal} {Nature}\ }\textbf {\bibinfo {volume} {571}},\ \bibinfo
  {pages} {371} (\bibinfo {year} {2019})}\BibitemShut {NoStop}%
\bibitem [{\citenamefont {Watson}\ \emph {et~al.}(2014)\citenamefont {Watson},
  \citenamefont {Weber}, \citenamefont {Miwa}, \citenamefont {Mahapatra},
  \citenamefont {Heijnen},\ and\ \citenamefont {Simmons}}]{Watson2014}%
  \BibitemOpen
  \bibfield  {author} {\bibinfo {author} {\bibfnamefont {T.~F.}\ \bibnamefont
  {Watson}}, \bibinfo {author} {\bibfnamefont {B.}~\bibnamefont {Weber}},
  \bibinfo {author} {\bibfnamefont {J.~A.}\ \bibnamefont {Miwa}}, \bibinfo
  {author} {\bibfnamefont {S.}~\bibnamefont {Mahapatra}}, \bibinfo {author}
  {\bibfnamefont {R.~M.~P.}\ \bibnamefont {Heijnen}},\ and\ \bibinfo {author}
  {\bibfnamefont {M.~Y.}\ \bibnamefont {Simmons}},\ }\bibfield  {title}
  {\bibinfo {title} {Transport in asymmetrically coupled donor-based silicon
  triple quantum dots},\ }\href {https://doi.org/10.1021/nl4045026} {\bibfield
  {journal} {\bibinfo  {journal} {Nano Lett.}\ }\textbf {\bibinfo {volume}
  {14}},\ \bibinfo {pages} {1830} (\bibinfo {year} {2014})}\BibitemShut
  {NoStop}%
\bibitem [{\citenamefont {Pakkiam}\ \emph {et~al.}(2018)\citenamefont
  {Pakkiam}, \citenamefont {Timofeev}, \citenamefont {House}, \citenamefont
  {Hogg}, \citenamefont {Kobayashi}, \citenamefont {Koch}, \citenamefont
  {Rogge},\ and\ \citenamefont {Simmons}}]{PrasannaPRX18}%
  \BibitemOpen
  \bibfield  {author} {\bibinfo {author} {\bibfnamefont {P.}~\bibnamefont
  {Pakkiam}}, \bibinfo {author} {\bibfnamefont {A.~V.}\ \bibnamefont
  {Timofeev}}, \bibinfo {author} {\bibfnamefont {M.~G.}\ \bibnamefont {House}},
  \bibinfo {author} {\bibfnamefont {M.~R.}\ \bibnamefont {Hogg}}, \bibinfo
  {author} {\bibfnamefont {T.}~\bibnamefont {Kobayashi}}, \bibinfo {author}
  {\bibfnamefont {M.}~\bibnamefont {Koch}}, \bibinfo {author} {\bibfnamefont
  {S.}~\bibnamefont {Rogge}},\ and\ \bibinfo {author} {\bibfnamefont {M.~Y.}\
  \bibnamefont {Simmons}},\ }\bibfield  {title} {\bibinfo {title} {Single-shot
  single-gate rf spin readout in silicon},\ }\href@noop {} {\bibfield
  {journal} {\bibinfo  {journal} {Phys. Rev. X}\ }\textbf {\bibinfo {volume}
  {8}},\ \bibinfo {pages} {041032} (\bibinfo {year} {2018})}\BibitemShut
  {NoStop}%
\bibitem [{\citenamefont {Hile}\ \emph {et~al.}(2018)\citenamefont {Hile},
  \citenamefont {Fricke}, \citenamefont {House}, \citenamefont {Peretz},
  \citenamefont {Chen}, \citenamefont {Wang}, \citenamefont {Broome},
  \citenamefont {Gorman}, \citenamefont {Keizer}, \citenamefont {Rahman},\ and\
  \citenamefont {Simmons}}]{Hile2018}%
  \BibitemOpen
  \bibfield  {author} {\bibinfo {author} {\bibfnamefont {S.~J.}\ \bibnamefont
  {Hile}}, \bibinfo {author} {\bibfnamefont {L.}~\bibnamefont {Fricke}},
  \bibinfo {author} {\bibfnamefont {M.~G.}\ \bibnamefont {House}}, \bibinfo
  {author} {\bibfnamefont {E.}~\bibnamefont {Peretz}}, \bibinfo {author}
  {\bibfnamefont {C.~Y.}\ \bibnamefont {Chen}}, \bibinfo {author}
  {\bibfnamefont {Y.}~\bibnamefont {Wang}}, \bibinfo {author} {\bibfnamefont
  {M.}~\bibnamefont {Broome}}, \bibinfo {author} {\bibfnamefont {S.~K.}\
  \bibnamefont {Gorman}}, \bibinfo {author} {\bibfnamefont {J.~G.}\
  \bibnamefont {Keizer}}, \bibinfo {author} {\bibfnamefont {R.}~\bibnamefont
  {Rahman}},\ and\ \bibinfo {author} {\bibfnamefont {M.~Y.}\ \bibnamefont
  {Simmons}},\ }\bibfield  {title} {\bibinfo {title} {Addressable electron spin
  resonance using donors and donor molecules in silicon},\ }\href
  {https://doi.org/10.1126/sciadv.aaq1459} {\bibfield  {journal} {\bibinfo
  {journal} {Science Advances}\ }\textbf {\bibinfo {volume} {4}},\ \bibinfo
  {pages} {eaaq1459} (\bibinfo {year} {2018})}\BibitemShut {NoStop}%
\bibitem [{\citenamefont {Tosi}\ \emph {et~al.}(2017)\citenamefont {Tosi},
  \citenamefont {Mohiyaddin}, \citenamefont {Schmitt}, \citenamefont {Tenberg},
  \citenamefont {Rahman}, \citenamefont {Klimeck},\ and\ \citenamefont
  {Morello}}]{Tosi2017}%
  \BibitemOpen
  \bibfield  {author} {\bibinfo {author} {\bibfnamefont {G.}~\bibnamefont
  {Tosi}}, \bibinfo {author} {\bibfnamefont {F.~A.}\ \bibnamefont
  {Mohiyaddin}}, \bibinfo {author} {\bibfnamefont {V.}~\bibnamefont {Schmitt}},
  \bibinfo {author} {\bibfnamefont {S.}~\bibnamefont {Tenberg}}, \bibinfo
  {author} {\bibfnamefont {R.}~\bibnamefont {Rahman}}, \bibinfo {author}
  {\bibfnamefont {G.}~\bibnamefont {Klimeck}},\ and\ \bibinfo {author}
  {\bibfnamefont {A.}~\bibnamefont {Morello}},\ }\bibfield  {title} {\bibinfo
  {title} {Silicon quantum processor with robust long-distance qubit
  couplings},\ }\href {https://doi.org/10.1038/s41467-017-00378-x} {\bibfield
  {journal} {\bibinfo  {journal} {Nature Communications}\ }\textbf {\bibinfo
  {volume} {8}},\ \bibinfo {pages} {450} (\bibinfo {year} {2017})}\BibitemShut
  {NoStop}%
\bibitem [{\citenamefont {Weber}\ \emph {et~al.}(2018)\citenamefont {Weber},
  \citenamefont {Hsueh}, \citenamefont {Watson}, \citenamefont {Li},
  \citenamefont {Hamilton}, \citenamefont {Hollenberg}, \citenamefont
  {Rahman},\ and\ \citenamefont {Simmons}}]{weber2018spin}%
  \BibitemOpen
  \bibfield  {author} {\bibinfo {author} {\bibfnamefont {B.}~\bibnamefont
  {Weber}}, \bibinfo {author} {\bibfnamefont {Y.-L.}\ \bibnamefont {Hsueh}},
  \bibinfo {author} {\bibfnamefont {T.~F.}\ \bibnamefont {Watson}}, \bibinfo
  {author} {\bibfnamefont {R.}~\bibnamefont {Li}}, \bibinfo {author}
  {\bibfnamefont {A.~R.}\ \bibnamefont {Hamilton}}, \bibinfo {author}
  {\bibfnamefont {L.~C.~L.}\ \bibnamefont {Hollenberg}}, \bibinfo {author}
  {\bibfnamefont {R.}~\bibnamefont {Rahman}},\ and\ \bibinfo {author}
  {\bibfnamefont {M.~Y.}\ \bibnamefont {Simmons}},\ }\bibfield  {title}
  {\bibinfo {title} {Spin--orbit coupling in silicon for electrons bound to
  donors},\ }\href@noop {} {\bibfield  {journal} {\bibinfo  {journal} {npj
  Quantum Information}\ }\textbf {\bibinfo {volume} {4}},\ \bibinfo {pages}
  {61} (\bibinfo {year} {2018})}\BibitemShut {NoStop}%
\bibitem [{\citenamefont {Ahmed*}\ \emph {et~al.}(2009)\citenamefont {Ahmed*},
  \citenamefont {Kharche*}, \citenamefont {Rahman*}, \citenamefont {Usman*},
  \citenamefont {Lee*}, \citenamefont {Ryu}, \citenamefont {Bae}, \citenamefont
  {Clark}, \citenamefont {Haley}, \citenamefont {Naumov}, \citenamefont
  {Saied}, \citenamefont {Korkusinski}, \citenamefont {Kennel}, \citenamefont
  {McLennan}, \citenamefont {Boykin},\ and\ \citenamefont
  {Klimeck}}]{Ahmed2009}%
  \BibitemOpen
  \bibfield  {author} {\bibinfo {author} {\bibfnamefont {S.}~\bibnamefont
  {Ahmed*}}, \bibinfo {author} {\bibfnamefont {N.}~\bibnamefont {Kharche*}},
  \bibinfo {author} {\bibfnamefont {R.}~\bibnamefont {Rahman*}}, \bibinfo
  {author} {\bibfnamefont {M.}~\bibnamefont {Usman*}}, \bibinfo {author}
  {\bibfnamefont {S.}~\bibnamefont {Lee*}}, \bibinfo {author} {\bibfnamefont
  {H.}~\bibnamefont {Ryu}}, \bibinfo {author} {\bibfnamefont {H.}~\bibnamefont
  {Bae}}, \bibinfo {author} {\bibfnamefont {S.}~\bibnamefont {Clark}}, \bibinfo
  {author} {\bibfnamefont {B.}~\bibnamefont {Haley}}, \bibinfo {author}
  {\bibfnamefont {M.}~\bibnamefont {Naumov}}, \bibinfo {author} {\bibfnamefont
  {F.}~\bibnamefont {Saied}}, \bibinfo {author} {\bibfnamefont
  {M.}~\bibnamefont {Korkusinski}}, \bibinfo {author} {\bibfnamefont
  {R.}~\bibnamefont {Kennel}}, \bibinfo {author} {\bibfnamefont
  {M.}~\bibnamefont {McLennan}}, \bibinfo {author} {\bibfnamefont {T.~B.}\
  \bibnamefont {Boykin}},\ and\ \bibinfo {author} {\bibfnamefont
  {G.}~\bibnamefont {Klimeck}},\ }\bibinfo {title} {Multimillion {A}tom
  {S}imulations with {Nemo3D}},\ in\ \href
  {https://doi.org/10.1007/978-0-387-30440-3_343} {\emph {\bibinfo {booktitle}
  {Encyclopedia of Complexity and Systems Science}}},\ \bibinfo {editor}
  {edited by\ \bibinfo {editor} {\bibfnamefont {R.~A.}\ \bibnamefont
  {Meyers}}}\ (\bibinfo  {publisher} {Springer New York},\ \bibinfo {address}
  {New York, NY},\ \bibinfo {year} {2009})\ pp.\ \bibinfo {pages}
  {5745--5783}\BibitemShut {NoStop}%
\bibitem [{\citenamefont {Benito}\ \emph {et~al.}(2017)\citenamefont {Benito},
  \citenamefont {Mi}, \citenamefont {Taylor}, \citenamefont {Petta},\ and\
  \citenamefont {Burkard}}]{Benito2017}%
  \BibitemOpen
  \bibfield  {author} {\bibinfo {author} {\bibfnamefont {M.}~\bibnamefont
  {Benito}}, \bibinfo {author} {\bibfnamefont {X.}~\bibnamefont {Mi}}, \bibinfo
  {author} {\bibfnamefont {J.~M.}\ \bibnamefont {Taylor}}, \bibinfo {author}
  {\bibfnamefont {J.~R.}\ \bibnamefont {Petta}},\ and\ \bibinfo {author}
  {\bibfnamefont {G.}~\bibnamefont {Burkard}},\ }\bibfield  {title} {\bibinfo
  {title} {Input-output theory for spin-photon coupling in {Si} double quantum
  dots},\ }\href {https://doi.org/10.1103/PhysRevB.96.235434} {\bibfield
  {journal} {\bibinfo  {journal} {Phys. Rev. B}\ }\textbf {\bibinfo {volume}
  {96}},\ \bibinfo {pages} {235434} (\bibinfo {year} {2017})}\BibitemShut
  {NoStop}%
\bibitem [{\citenamefont {Boross}\ \emph {et~al.}(2018)\citenamefont {Boross},
  \citenamefont {Sz\'echenyi},\ and\ \citenamefont {P\'alyi}}]{Boross2018}%
  \BibitemOpen
  \bibfield  {author} {\bibinfo {author} {\bibfnamefont {P.}~\bibnamefont
  {Boross}}, \bibinfo {author} {\bibfnamefont {G.}~\bibnamefont
  {Sz\'echenyi}},\ and\ \bibinfo {author} {\bibfnamefont {A.}~\bibnamefont
  {P\'alyi}},\ }\bibfield  {title} {\bibinfo {title} {Hyperfine-assisted fast
  electric control of dopant nuclear spins in semiconductors},\ }\href
  {https://doi.org/10.1103/PhysRevB.97.245417} {\bibfield  {journal} {\bibinfo
  {journal} {Phys. Rev. B}\ }\textbf {\bibinfo {volume} {97}},\ \bibinfo
  {pages} {245417} (\bibinfo {year} {2018})}\BibitemShut {NoStop}%
\bibitem [{\citenamefont {Het\'enyi}\ \emph {et~al.}(2019)\citenamefont
  {Het\'enyi}, \citenamefont {Boross},\ and\ \citenamefont
  {P\'alyi}}]{Hetenyi2019}%
  \BibitemOpen
  \bibfield  {author} {\bibinfo {author} {\bibfnamefont {B.}~\bibnamefont
  {Het\'enyi}}, \bibinfo {author} {\bibfnamefont {P.}~\bibnamefont {Boross}},\
  and\ \bibinfo {author} {\bibfnamefont {A.}~\bibnamefont {P\'alyi}},\
  }\bibfield  {title} {\bibinfo {title} {Hyperfine-assisted decoherence of a
  phosphorus nuclear-spin qubit in silicon},\ }\href
  {https://doi.org/10.1103/PhysRevB.100.115435} {\bibfield  {journal} {\bibinfo
   {journal} {Phys. Rev. B}\ }\textbf {\bibinfo {volume} {100}},\ \bibinfo
  {pages} {115435} (\bibinfo {year} {2019})}\BibitemShut {NoStop}%
\bibitem [{\citenamefont {Klimeck}\ \emph {et~al.}(2002)\citenamefont
  {Klimeck}, \citenamefont {Oyafuso}, \citenamefont {Boykin}, \citenamefont
  {Bowen},\ and\ \citenamefont {von Allmen}}]{nemo_ref_1}%
  \BibitemOpen
  \bibfield  {author} {\bibinfo {author} {\bibfnamefont {G.}~\bibnamefont
  {Klimeck}}, \bibinfo {author} {\bibfnamefont {F.}~\bibnamefont {Oyafuso}},
  \bibinfo {author} {\bibfnamefont {T.~B.}\ \bibnamefont {Boykin}}, \bibinfo
  {author} {\bibfnamefont {R.~C.}\ \bibnamefont {Bowen}},\ and\ \bibinfo
  {author} {\bibfnamefont {P.}~\bibnamefont {von Allmen}},\ }\bibfield  {title}
  {\bibinfo {title} {Development of a {N}anoelectronic {3-D} ({NEMO 3-D} )
  {S}imulator for {M}ultimillion {A}tom {S}imulations and {I}ts {A}pplication
  to {A}lloyed {Q}uantum {D}ots},\ }\href
  {https://doi.org/10.3970/cmes.2002.003.601} {\bibfield  {journal} {\bibinfo
  {journal} {Computer Modeling in Engineering \& Sciences}\ }\textbf {\bibinfo
  {volume} {3}},\ \bibinfo {pages} {601} (\bibinfo {year} {2002})}\BibitemShut
  {NoStop}%
\bibitem [{\citenamefont {{Klimeck}}\ \emph {et~al.}(2007)\citenamefont
  {{Klimeck}}, \citenamefont {{Ahmed}}, \citenamefont {{Hansang Bae}},
  \citenamefont {{Kharche}}, \citenamefont {{Clark}}, \citenamefont {{Haley}},
  \citenamefont {{Sunhee Lee}}, \citenamefont {{Naumov}}, \citenamefont {{Hoon
  Ryu}}, \citenamefont {{Saied}}, \citenamefont {{Prada}}, \citenamefont
  {{Korkusinski}}, \citenamefont {{Boykin}},\ and\ \citenamefont
  {{Rahman}}}]{nemo_ref_2}%
  \BibitemOpen
  \bibfield  {author} {\bibinfo {author} {\bibfnamefont {G.}~\bibnamefont
  {{Klimeck}}}, \bibinfo {author} {\bibfnamefont {S.~S.}\ \bibnamefont
  {{Ahmed}}}, \bibinfo {author} {\bibnamefont {{Hansang Bae}}}, \bibinfo
  {author} {\bibfnamefont {N.}~\bibnamefont {{Kharche}}}, \bibinfo {author}
  {\bibfnamefont {S.}~\bibnamefont {{Clark}}}, \bibinfo {author} {\bibfnamefont
  {B.}~\bibnamefont {{Haley}}}, \bibinfo {author} {\bibnamefont {{Sunhee
  Lee}}}, \bibinfo {author} {\bibfnamefont {M.}~\bibnamefont {{Naumov}}},
  \bibinfo {author} {\bibnamefont {{Hoon Ryu}}}, \bibinfo {author}
  {\bibfnamefont {F.}~\bibnamefont {{Saied}}}, \bibinfo {author} {\bibfnamefont
  {M.}~\bibnamefont {{Prada}}}, \bibinfo {author} {\bibfnamefont
  {M.}~\bibnamefont {{Korkusinski}}}, \bibinfo {author} {\bibfnamefont {T.~B.}\
  \bibnamefont {{Boykin}}},\ and\ \bibinfo {author} {\bibfnamefont
  {R.}~\bibnamefont {{Rahman}}},\ }\bibfield  {title} {\bibinfo {title}
  {Atomistic simulation of realistically sized nanodevices using nemo
  3-d—part i: Models and benchmarks},\ }\href
  {https://doi.org/10.1109/TED.2007.902879} {\bibfield  {journal} {\bibinfo
  {journal} {IEEE Transactions on Electron Devices}\ }\textbf {\bibinfo
  {volume} {54}},\ \bibinfo {pages} {2079} (\bibinfo {year}
  {2007})}\BibitemShut {NoStop}%
\bibitem [{\citenamefont {Wang}\ \emph {et~al.}(2016)\citenamefont {Wang},
  \citenamefont {Chen}, \citenamefont {Klimeck}, \citenamefont {Simmons},\ and\
  \citenamefont {Rahman}}]{Wang2016}%
  \BibitemOpen
  \bibfield  {author} {\bibinfo {author} {\bibfnamefont {Y.}~\bibnamefont
  {Wang}}, \bibinfo {author} {\bibfnamefont {C.-Y.}\ \bibnamefont {Chen}},
  \bibinfo {author} {\bibfnamefont {G.}~\bibnamefont {Klimeck}}, \bibinfo
  {author} {\bibfnamefont {M.~Y.}\ \bibnamefont {Simmons}},\ and\ \bibinfo
  {author} {\bibfnamefont {R.}~\bibnamefont {Rahman}},\ }\bibfield  {title}
  {\bibinfo {title} {Characterizing {S}i:{P} quantum dot qubits with spin
  resonance techniques},\ }\href {https://doi.org/10.1038/srep31830} {\bibfield
   {journal} {\bibinfo  {journal} {Scientific Reports}\ }\textbf {\bibinfo
  {volume} {6}},\ \bibinfo {pages} {31830} (\bibinfo {year}
  {2016})}\BibitemShut {NoStop}%
\bibitem [{\citenamefont {Pla}\ \emph {et~al.}(2013)\citenamefont {Pla},
  \citenamefont {Tan}, \citenamefont {Dehollain}, \citenamefont {Lim},
  \citenamefont {Morton}, \citenamefont {Zwanenburg}, \citenamefont {Jamieson},
  \citenamefont {Dzurak},\ and\ \citenamefont {Morello}}]{Pla2013}%
  \BibitemOpen
  \bibfield  {author} {\bibinfo {author} {\bibfnamefont {J.~J.}\ \bibnamefont
  {Pla}}, \bibinfo {author} {\bibfnamefont {K.~Y.}\ \bibnamefont {Tan}},
  \bibinfo {author} {\bibfnamefont {J.~P.}\ \bibnamefont {Dehollain}}, \bibinfo
  {author} {\bibfnamefont {W.~H.}\ \bibnamefont {Lim}}, \bibinfo {author}
  {\bibfnamefont {J.~J.~L.}\ \bibnamefont {Morton}}, \bibinfo {author}
  {\bibfnamefont {F.~A.}\ \bibnamefont {Zwanenburg}}, \bibinfo {author}
  {\bibfnamefont {D.~N.}\ \bibnamefont {Jamieson}}, \bibinfo {author}
  {\bibfnamefont {A.~S.}\ \bibnamefont {Dzurak}},\ and\ \bibinfo {author}
  {\bibfnamefont {A.}~\bibnamefont {Morello}},\ }\bibfield  {title} {\bibinfo
  {title} {High-fidelity readout and control of a nuclear spin qubit in
  silicon},\ }\href {https://doi.org/10.1038/nature12011} {\bibfield  {journal}
  {\bibinfo  {journal} {Nature}\ }\textbf {\bibinfo {volume} {496}},\ \bibinfo
  {pages} {334} (\bibinfo {year} {2013})}\BibitemShut {NoStop}%
\bibitem [{\citenamefont {Abragam}\ and\ \citenamefont
  {Goldman}(1978)}]{Abragam1978}%
  \BibitemOpen
  \bibfield  {author} {\bibinfo {author} {\bibfnamefont {A.}~\bibnamefont
  {Abragam}}\ and\ \bibinfo {author} {\bibfnamefont {M.}~\bibnamefont
  {Goldman}},\ }\bibfield  {title} {\bibinfo {title} {Principles of dynamic
  nuclear polarisation},\ }\href {https://doi.org/10.1088/0034-4885/41/3/002}
  {\bibfield  {journal} {\bibinfo  {journal} {Reports on Progress in Physics}\
  }\textbf {\bibinfo {volume} {41}},\ \bibinfo {pages} {395} (\bibinfo {year}
  {1978})}\BibitemShut {NoStop}%
\bibitem [{\citenamefont {Simmons}\ \emph {et~al.}(2011)\citenamefont
  {Simmons}, \citenamefont {Brown}, \citenamefont {Riemann}, \citenamefont
  {Abrosimov}, \citenamefont {Becker}, \citenamefont {Pohl}, \citenamefont
  {Thewalt}, \citenamefont {Itoh},\ and\ \citenamefont {Morton}}]{Simmons2011}%
  \BibitemOpen
  \bibfield  {author} {\bibinfo {author} {\bibfnamefont {S.}~\bibnamefont
  {Simmons}}, \bibinfo {author} {\bibfnamefont {R.~M.}\ \bibnamefont {Brown}},
  \bibinfo {author} {\bibfnamefont {H.}~\bibnamefont {Riemann}}, \bibinfo
  {author} {\bibfnamefont {N.~V.}\ \bibnamefont {Abrosimov}}, \bibinfo {author}
  {\bibfnamefont {P.}~\bibnamefont {Becker}}, \bibinfo {author} {\bibfnamefont
  {H.-J.}\ \bibnamefont {Pohl}}, \bibinfo {author} {\bibfnamefont {M.~L.~W.}\
  \bibnamefont {Thewalt}}, \bibinfo {author} {\bibfnamefont {K.~M.}\
  \bibnamefont {Itoh}},\ and\ \bibinfo {author} {\bibfnamefont {J.~J.~L.}\
  \bibnamefont {Morton}},\ }\bibfield  {title} {\bibinfo {title} {Entanglement
  in a solid-state spin ensemble},\ }\href
  {https://doi.org/10.1038/nature09696} {\bibfield  {journal} {\bibinfo
  {journal} {Nature}\ }\textbf {\bibinfo {volume} {470}},\ \bibinfo {pages}
  {69} (\bibinfo {year} {2011})}\BibitemShut {NoStop}%
\bibitem [{\citenamefont {Yoneda}\ \emph {et~al.}(2018)\citenamefont {Yoneda},
  \citenamefont {Takeda}, \citenamefont {Otsuka}, \citenamefont {Nakajima},
  \citenamefont {Delbecq}, \citenamefont {Allison}, \citenamefont {Honda},
  \citenamefont {Kodera}, \citenamefont {Oda}, \citenamefont {Hoshi},
  \citenamefont {Usami}, \citenamefont {Itoh},\ and\ \citenamefont
  {Tarucha}}]{Yoneda2018}%
  \BibitemOpen
  \bibfield  {author} {\bibinfo {author} {\bibfnamefont {J.}~\bibnamefont
  {Yoneda}}, \bibinfo {author} {\bibfnamefont {K.}~\bibnamefont {Takeda}},
  \bibinfo {author} {\bibfnamefont {T.}~\bibnamefont {Otsuka}}, \bibinfo
  {author} {\bibfnamefont {T.}~\bibnamefont {Nakajima}}, \bibinfo {author}
  {\bibfnamefont {M.~R.}\ \bibnamefont {Delbecq}}, \bibinfo {author}
  {\bibfnamefont {G.}~\bibnamefont {Allison}}, \bibinfo {author} {\bibfnamefont
  {T.}~\bibnamefont {Honda}}, \bibinfo {author} {\bibfnamefont
  {T.}~\bibnamefont {Kodera}}, \bibinfo {author} {\bibfnamefont
  {S.}~\bibnamefont {Oda}}, \bibinfo {author} {\bibfnamefont {Y.}~\bibnamefont
  {Hoshi}}, \bibinfo {author} {\bibfnamefont {N.}~\bibnamefont {Usami}},
  \bibinfo {author} {\bibfnamefont {K.~M.}\ \bibnamefont {Itoh}},\ and\
  \bibinfo {author} {\bibfnamefont {S.}~\bibnamefont {Tarucha}},\ }\bibfield
  {title} {\bibinfo {title} {A quantum-dot spin qubit with coherence limited by
  charge noise and fidelity higher than 99.9\%},\ }\href
  {https://doi.org/10.1038/s41565-017-0014-x} {\bibfield  {journal} {\bibinfo
  {journal} {Nature Nanotechnology}\ }\textbf {\bibinfo {volume} {13}},\
  \bibinfo {pages} {102} (\bibinfo {year} {2018})}\BibitemShut {NoStop}%
\bibitem [{\citenamefont {Kranz}\ \emph {et~al.}(2020)\citenamefont {Kranz},
  \citenamefont {Gorman}, \citenamefont {Thorgrimsson}, \citenamefont {He},
  \citenamefont {Keith}, \citenamefont {Keizer},\ and\ \citenamefont
  {Simmons}}]{Kranz2020}%
  \BibitemOpen
  \bibfield  {author} {\bibinfo {author} {\bibfnamefont {L.}~\bibnamefont
  {Kranz}}, \bibinfo {author} {\bibfnamefont {S.~K.}\ \bibnamefont {Gorman}},
  \bibinfo {author} {\bibfnamefont {B.}~\bibnamefont {Thorgrimsson}}, \bibinfo
  {author} {\bibfnamefont {Y.}~\bibnamefont {He}}, \bibinfo {author}
  {\bibfnamefont {D.}~\bibnamefont {Keith}}, \bibinfo {author} {\bibfnamefont
  {J.~G.}\ \bibnamefont {Keizer}},\ and\ \bibinfo {author} {\bibfnamefont
  {M.~Y.}\ \bibnamefont {Simmons}},\ }\bibfield  {title} {\bibinfo {title}
  {Exploiting a single-crystal environment to minimize the charge noise on
  qubits in silicon},\ }\href {https://doi.org/10.1002/adma.202003361}
  {\bibfield  {journal} {\bibinfo  {journal} {Advanced Materials}\ }\textbf
  {\bibinfo {volume} {32}},\ \bibinfo {pages} {2003361} (\bibinfo {year}
  {2020})}\BibitemShut {NoStop}%
\bibitem [{\citenamefont {Struck}\ \emph {et~al.}(2020)\citenamefont {Struck},
  \citenamefont {Hollmann}, \citenamefont {Schauer}, \citenamefont {Fedorets},
  \citenamefont {Schmidbauer}, \citenamefont {Sawano}, \citenamefont {Riemann},
  \citenamefont {Abrosimov}, \citenamefont {Cywi{\'{n}}ski}, \citenamefont
  {Bougeard},\ and\ \citenamefont {Schreiber}}]{SchreiberNJP20}%
  \BibitemOpen
  \bibfield  {author} {\bibinfo {author} {\bibfnamefont {T.}~\bibnamefont
  {Struck}}, \bibinfo {author} {\bibfnamefont {A.}~\bibnamefont {Hollmann}},
  \bibinfo {author} {\bibfnamefont {F.}~\bibnamefont {Schauer}}, \bibinfo
  {author} {\bibfnamefont {O.}~\bibnamefont {Fedorets}}, \bibinfo {author}
  {\bibfnamefont {A.}~\bibnamefont {Schmidbauer}}, \bibinfo {author}
  {\bibfnamefont {K.}~\bibnamefont {Sawano}}, \bibinfo {author} {\bibfnamefont
  {H.}~\bibnamefont {Riemann}}, \bibinfo {author} {\bibfnamefont {N.~V.}\
  \bibnamefont {Abrosimov}}, \bibinfo {author} {\bibfnamefont
  {{\L}.}~\bibnamefont {Cywi{\'{n}}ski}}, \bibinfo {author} {\bibfnamefont
  {D.}~\bibnamefont {Bougeard}},\ and\ \bibinfo {author} {\bibfnamefont
  {L.~R.}\ \bibnamefont {Schreiber}},\ }\bibfield  {title} {\bibinfo {title}
  {Low-frequency spin qubit energy splitting noise in highly purified
  {$^{28}$}{Si/SiGe}},\ }\href {https://doi.org/10.1038/s41534-020-0276-2}
  {\bibfield  {journal} {\bibinfo  {journal} {npj Quantum Information}\
  }\textbf {\bibinfo {volume} {6}},\ \bibinfo {pages} {40} (\bibinfo {year}
  {2020})}\BibitemShut {NoStop}%
\bibitem [{\citenamefont {Ruess}\ \emph {et~al.}(2004)\citenamefont {Ruess},
  \citenamefont {Oberbeck}, \citenamefont {Simmons}, \citenamefont {Goh},
  \citenamefont {Hamilton}, \citenamefont {Hallam}, \citenamefont {Schofield},
  \citenamefont {Curson},\ and\ \citenamefont {Clark}}]{Ruess2004}%
  \BibitemOpen
  \bibfield  {author} {\bibinfo {author} {\bibfnamefont {F.~J.}\ \bibnamefont
  {Ruess}}, \bibinfo {author} {\bibfnamefont {L.}~\bibnamefont {Oberbeck}},
  \bibinfo {author} {\bibfnamefont {M.~Y.}\ \bibnamefont {Simmons}}, \bibinfo
  {author} {\bibfnamefont {K.~E.~J.}\ \bibnamefont {Goh}}, \bibinfo {author}
  {\bibfnamefont {A.~R.}\ \bibnamefont {Hamilton}}, \bibinfo {author}
  {\bibfnamefont {T.}~\bibnamefont {Hallam}}, \bibinfo {author} {\bibfnamefont
  {S.~R.}\ \bibnamefont {Schofield}}, \bibinfo {author} {\bibfnamefont {N.~J.}\
  \bibnamefont {Curson}},\ and\ \bibinfo {author} {\bibfnamefont {R.~G.}\
  \bibnamefont {Clark}},\ }\bibfield  {title} {\bibinfo {title} {Toward
  atomic-scale device fabrication in silicon using scanning probe microscopy},\
  }\href@noop {} {\bibfield  {journal} {\bibinfo  {journal} {Nano Letters}\
  }\textbf {\bibinfo {volume} {4}},\ \bibinfo {pages} {1969} (\bibinfo {year}
  {2004})}\BibitemShut {NoStop}%
\bibitem [{\citenamefont {Samkharadze}\ \emph {et~al.}(2016)\citenamefont
  {Samkharadze}, \citenamefont {Bruno}, \citenamefont {Scarlino}, \citenamefont
  {Zheng}, \citenamefont {DiVincenzo}, \citenamefont {DiCarlo},\ and\
  \citenamefont {Vandersypen}}]{VandersypenPhysRevAppl19}%
  \BibitemOpen
  \bibfield  {author} {\bibinfo {author} {\bibfnamefont {N.}~\bibnamefont
  {Samkharadze}}, \bibinfo {author} {\bibfnamefont {A.}~\bibnamefont {Bruno}},
  \bibinfo {author} {\bibfnamefont {P.}~\bibnamefont {Scarlino}}, \bibinfo
  {author} {\bibfnamefont {G.}~\bibnamefont {Zheng}}, \bibinfo {author}
  {\bibfnamefont {D.~P.}\ \bibnamefont {DiVincenzo}}, \bibinfo {author}
  {\bibfnamefont {L.}~\bibnamefont {DiCarlo}},\ and\ \bibinfo {author}
  {\bibfnamefont {L.~M.~K.}\ \bibnamefont {Vandersypen}},\ }\bibfield  {title}
  {\bibinfo {title} {High-kinetic-inductance superconducting nanowire
  resonators for circuit {QED} in a magnetic field},\ }\href@noop {} {\bibfield
   {journal} {\bibinfo  {journal} {Phys. Rev. Applied}\ }\textbf {\bibinfo
  {volume} {5}},\ \bibinfo {pages} {044004} (\bibinfo {year}
  {2016})}\BibitemShut {NoStop}%
\bibitem [{\citenamefont {Probst}\ \emph {et~al.}(2015)\citenamefont {Probst},
  \citenamefont {Song}, \citenamefont {Bushev}, \citenamefont {Ustinov},\ and\
  \citenamefont {Weides}}]{Probst15}%
  \BibitemOpen
  \bibfield  {author} {\bibinfo {author} {\bibfnamefont {S.}~\bibnamefont
  {Probst}}, \bibinfo {author} {\bibfnamefont {F.~B.}\ \bibnamefont {Song}},
  \bibinfo {author} {\bibfnamefont {P.~A.}\ \bibnamefont {Bushev}}, \bibinfo
  {author} {\bibfnamefont {A.~V.}\ \bibnamefont {Ustinov}},\ and\ \bibinfo
  {author} {\bibfnamefont {M.}~\bibnamefont {Weides}},\ }\bibfield  {title}
  {\bibinfo {title} {Efficient and robust analysis of complex scattering data
  under noise in microwave resonators},\ }\href@noop {} {\bibfield  {journal}
  {\bibinfo  {journal} {Rev. Sci. Instrum.}\ }\textbf {\bibinfo {volume}
  {86}},\ \bibinfo {pages} {024706} (\bibinfo {year} {2015})}\BibitemShut
  {NoStop}%
\end{thebibliography}%
	
\end{document}